\documentclass[useAMS,usenatbib]{mnras}

\usepackage[english]{babel}
\usepackage{amsmath,amssymb,bm}

\usepackage{graphicx}

\usepackage[normalem]{ulem}

\usepackage{verbatim}
\usepackage{color}
\usepackage{multirow}
\usepackage{mathtools}
\usepackage{epstopdf}

\usepackage{url}
\usepackage{xcolor}

\usepackage{gensymb}

\usepackage[usestackEOL]{stackengine}

\def\nat{Nature}

\def\mnras{MNRAS}
\def\apj{ApJ}
\def\apjl{ApJL}
\def\apjs{ApJS}
\def\aap{A\&A}

\def\aapr{A\&A Rev.}
\def\araa{ARA\&A}
\def\aj{AJ}

\def\pasp{Publ. Astr. Soc. Pac.}

\def\be{\begin{equation}} 
\def\ee{\end{equation}} 
\def\ba{\begin{eqnarray}} 
\def\ea{\end{eqnarray}}

\def\cc{\,{\rm {cm^{-3}}}} 
\def\msun{{\Msun}}

\def\gsim{\lower.5ex\hbox{\gtsima}} 
\def\lsim{\lower.5ex\hbox{\ltsima}} \def\gtsima{$\; \buildrel > \over 
\sim \;$} \def\ltsima{$\; \buildrel < \over \sim \;$} \def\prosima{$\; 
\buildrel \propto \over \sim \;$} \def\gsim{\lower.5ex\hbox{\gtsima}} 
\def\lsim{\lower.5ex\hbox{\ltsima}} 
\def\simgt{\lower.5ex\hbox{\gtsima}} 
\def\simlt{\lower.5ex\hbox{\ltsima}} 
\def\simpr{\lower.5ex\hbox{\prosima}} \def\la{\lsim} \def\ga{\gsim} 
  
 \def\gtsima{$\; \buildrel > \over \sim \;$} 
\def\ltsima{$\; \buildrel < \over \sim \;$} 
\def\gsim{\lower.5ex\hbox{\gtsima}} 
\def\lsim{\lower.5ex\hbox{\ltsima}} 
\def\simgt{\lower.5ex\hbox{\gtsima}} 
\def\simlt{\lower.5ex\hbox{\ltsima}} 
\def\simpr{\lower.5ex\hbox{\prosima}} 
\def\la{\lsim} 
\def\ga{\gsim}

\def\msun{\,{\rm \Msun}}

\def\E3{{\cal E}_{\rm g}^{III}}

\def\r12{r_{1/2}} 
\def\x12{x_{1/2}} 
\def\v12{v_{1/2}}

\def\gcc{\rm g \,{\rm {cm^{-3}}}}

\newcommand\code[1]{\textsc{\MakeLowercase{#1}}}

\def\nh2{n_{\rm H2}}
\def\fh2{f_{\rm H2}}

\def\mp{m_{\rm p}}

\def\angstrom{\textrm{A\kern -1.3ex\raisebox{0.6ex}{$^\circ$}}}

\def\mum{\mu{\rm m}}
\def\msun{{\rm M}_{\odot}}
\def\zsun{{\rm Z}_{\odot}}
\def\lsun{{\rm L}_{\odot}}

\def\cc{{\rm cm}^{-3}}

\def\msunyr{\msun\,{\rm yr}^{-1}}

\def\surfdpc{\msun\,{\rm pc}^{-2}}

\def\extccg{{\rm cm}^2\,{\rm g}^{-1}}

\def\SFR{{\rm SFR}}

\def\noAGN{\emph{noAGN}}

\def\AGNcone{\emph{AGNcone}}
\def\amin{a_{\rm min}}
\def\amax{a_{\rm max}}
\def\Rsmall{\mathcal{R_{\rm small}}}
\def\Rlarge{\mathcal{R_{\rm large}}}
\def\Rtotal{\mathcal{R_{\rm total}}}
\def\tauV{\tau_{\rm V}}

\defcitealias{paramita:2018}{B18}
\defcitealias{Gallerani:2010}{G10}
\defcitealias{DiMascia:2021}{DM21}

\graphicspath{{figures/RT_plots/AGNconeBHon008SMC/},{figures/RT_plots/AGNsphereBHon008SMC/},{figures/RT_plots/noAGN008SMC/},{figures/RT_plots/AGNconeBHon03SMC/},{figures/RT_plots/AGNsphereBHon03SMC/},{figures/RT_plots/noAGN03SMC/},{figures/RT_plots/noAGN008MW/},{figures/RT_plots/noAGN03MW/},{figures/RT_plots/AGNsphereBHon008MW/},{figures/RT_plots/AGNsphereBHon03MW/},{figures/RT_plots/AGNconeBHon008MW/},{figures/RT_plots/AGNconeBHon03MW/},{figures/comparisons/},{figures/extinction/},{figures/special_sims/}, {figures/chi_square/}, {figures/add_extinction/}}

\date{}
\pagerange{\pageref{firstpage}--\pageref{lastpage}} \pubyear{2021}

\title[The dust attenuation law in $z\sim 6$ quasars]{The dust attenuation law in $z\sim 6$ quasars}
\author[F. Di Mascia et al.]{F. Di Mascia$^{1}$\thanks{\href{mailto:fabio.dimascia@sns.it}{fabio.dimascia@sns.it}}, S. Gallerani$^{1}$, A. Ferrara$^{1}$, A. Pallottini$^{1}$, R. Maiolino$^{2,3,4}$ 
\newauthor S. Carniani$^{1}$, V. D'Odorico$^{1,5}$
\\
$^{1}$Scuola Normale Superiore, Piazza dei Cavalieri 7, I-56126 Pisa, Italy\\
$^{2}$Cavendish Laboratory, University of Cambridge, 19 J.J. Thomson Avenue, Cambridge CB3 0HE, UK\\
$^{3}$Kavli Institute for Cosmology, University of Cambridge, Madingley Road, Cambridge CB3 0HA, UK\\ 
$^{4}$Department of Physics and Astronomy, University College London, Gower Street, London WC1E 6BT, UK\\
$^{5}$INAF - Osservatorio Astronomico di Trieste, via G.B. Tiepolo, 11 I-34143 Trieste, Italy
}

\begin{document}

\maketitle
\label{firstpage}
\begin{abstract} 
We investigate the attenuation law in $z\sim 6$ quasars by combining cosmological zoom-in hydrodynamical simulations of quasar host galaxies, with multi-frequency radiative transfer calculations. We consider several dust models differing in terms of grain size distributions, dust mass and chemical composition, and compare the resulting synthetic Spectral Energy Distributions (SEDs) with data from bright, early quasars. We show that only dust models with grain size distributions in which small grains ($a\lesssim 0.1~\mum$, corresponding to $\approx 60\%$ of the total dust mass) are selectively removed from the dusty medium provide a good fit to the data. Removal can occur if small grains are efficiently destroyed in quasar environments and/or early dust production preferentially results in large grains. Attenuation  curves for these  models  are  close  to  flat, and consistent with recent data; they correspond to an effective dust-to-metal ratio $f_d \simeq 0.38$, i.e. close to the Milky Way value. 
\end{abstract}


\section{Introduction}

Dust is a key ingredient in understanding the properties of galaxies both from a theoretical and an observational point of view. Dust enhances the formation of molecular hydrogen, which is the main coolant in primordial gas, with an efficiency depending on the grains surfaces \citep{HirashitaFerrara:2002, CazauxTielens:2004}. Dust cooling also becomes relevant in the later stage of star formation, affecting the typical mass of stars and shaping the stellar initial mass function (IMF; \citealt{Larson:2005, Omukai:2005, Schneider:2006}). Therefore, both the overall dust content and the dust grain size distribution play an important role in the physical processes regulating star formation. 

Dust grains also scatter and absorb optical-ultraviolet (UV) radiation from stars and accreting BHs, and thermally re-emit this energy as infrared (IR) photons. Consequently dust affects the spectral energy distribution (SED) of observed galaxies \citep[e.g.][]{Calzetti:2000}, the intensity of emission lines and the estimate of the star-formation rate (SFR), which requires corrections for dust extinction \citep[e.g.][]{Kennicutt:2012}. Therefore, constraining dust properties is essential for interpreting observations and understanding galaxy evolution.

Dust is thought to form via condensation of metals in the atmosphere of evolved asymptotic giant branch (AGB) stars \citep[e.g.][]{Ferrarotti:2006} or in supernova (SN) ejecta \citep[SNe; e.g.][]{Kozasa:1989, Todini:2001, Nozawa:2003, Bianchi:2007}. However, how much mass is produced by each channel is still a matter of debate, especially at the high-redshift \citep{Morgan:2003, Valiante:2009, Gall2011, Leniewska2019}. 

The first grains formed in the Universe undergo several processes which alter the total dust mass and the initial grains size distribution. Grain-grain collisions in the ISM or in SN shocks convert large grains into small grains \citep[shattering,][]{Jones1996ApJ, Yan2004}, whereas small grains can accrete gas-phase metals in the dense ISM or in molecular clouds (MCs), increasing their size and the overall dust mass content (\citealt{Spitzer1978, Dwek1980ApJ, Draine2009ASPC}, but see also \citealt{Ferrara:2016b}) or growing into larger grains via coagulation \citep{Chokshi1993, Hirashita2009}. Gas-grains collisions (sputtering) in SN shocks or in hot plasma destroy dust grains and return their metals back to the gas \citep{Draine:1979, Tielens:1994}. The aforementioned processes are the main drivers of the dust-cycle in galaxies, which is deeply interconnected with their evolution. 

The wavelength dependence of dust extinction entangles information on the grain chemical composition and size distribution, and it is therefore a powerful observational tool to constrain dust models. Measurements of the extinction curve have been performed for several lines of sight in the Milky Way \citep[MW;][]{Savage:1979}, Small Magellanic Cloud (SMC) and Large Magellanic Cloud \citep[LMC;][]{Fitzpatrick:1989}. These observations are very well explained by a model in which the Milky Way dust is composed by carbonaceous and silicate grains, while the SMC extinction curve is dominated by silicates and the LMC represents an intermediate case \citep{Weingartner:2001}.

In general, inferring the dust extinction curve for star-forming galaxies is not trivial, because the observed spectrum also depends on the relative distribution of dust and stars. Therefore, observations of star-forming galaxies typically provide an \emph{attenuation} law, which describes the effective reddening of the radiation as compared to the case with no extinction. \citet{Calzetti:1994} found an attenuation law for local star-forming galaxies which is flatter than the extinction curve for SMC, LMC and MW. In particular, the resulting curve is characterised by the absence of the $2175$ $\angstrom$ dust feature, which is prominent in the MW curve and is usually attributed to polycyclic aromatic hydrocarbon (PAH) nanoparticles \citep[e.g.][]{Li:2007}. This result can be explained with an SMC-like intrinsic curve assuming a clumpy distribution for dust and stars \citep{Gordon:1997,Inoue:2005} or with a MW-type dust for different geometries \citep{Pierini:2004,Panuzzo:2007}. Interpreting observations of star-forming galaxies and inferring their dust properties is therefore quite difficult.

Active Galactic Nuclei (AGN), whose UV emission is dominated by the accretion disk powering the central black hole (BH) \citep[e.g.][]{kormendy2013}, represent an alternative place where to study dust extinction properties, possibly avoiding the complications presented by the dust/stars distribution in star-forming galaxies. Observations of mildly reddened quasars at $z<4.4$ suggested extinction curves compatible with the SMC one \citep{Reichard:2003, Richards2003AJ, Hopkins:2004}. Instead, highly obscured quasars show a prevalence of extinction curves markedly different from any of the SMC/LMC/MW ones, as indicated both by individual sources at $z<0.7$ \citep{Maiolino:2001a, Gaskell:2007}, and composite spectra (\citealt{Gaskell:2004, Czerny:2004}, but see also \citealt{Willott2005ApJ}). These works underlined two important features of extinction curves in AGN: (i) a flat (or \emph{grey}) extinction in the far-ultraviolet (FUV), and (ii) the absence of the $2175$ $\angstrom$ bump. A grey extinction was interpreted as an indication of the dominance of large grains in the circumnuclear region of AGN, possibly because of the depletion of small grains \citep{LaorDraine1993} or efficient coagulation (\citealt{Maiolino:2001b}; but see also \citealt{Weingartner2002ApJ}).

The picture is even more complex at high-redshift. \citet{Maiolino:2004} analysed the spectrum of the reddened broad absorption line (BAL) quasar SDSSJ1048+4637 at $z=6.2$, revealing a peculiar extinction curve, being flat at rest-frame $\lambda \gtrsim 1700~\angstrom$, and rising at $\lambda \lesssim 1700~\angstrom$, compatible with a supernova origin \citep{Todini:2001, Hirashita2005, Bianchi:2007}. A similar extinction curve was also inferred for the quasar CFHQS J1509-1749 at $z=6.12$ \citep{Willott2007CFHQS}. \citet[][hereafter {\citetalias{Gallerani:2010}}]{Gallerani:2010} studied the optical-near infrared spectra of 33 quasars at $3.9<z<6.4$, finding that the extinction curve required to explain dust-reddened quasars (characterised by an attenuation at $3000~\angstrom$, $A_{3000}$, in the range $0.82 < A_{3000} < 2.0$) deviate from the SMC extinction curve, with a tendency to flatten at $\lambda \lesssim 2000~\angstrom$. In the case of BAL quasars, they also computed a mean extinction curve \citepalias[MECBAL in][]{Gallerani:2010} which displays a significant flattening at $\lambda \lesssim 2000~\angstrom$. These observations suggest that dust properties in AGN at high-redshift might be quite different from those deduced in the local Universe \citep[but see also][]{Hjorth2013ApJ}.  

A different dust origin at high-redshift is also supported by the timescales involved in dust production mechanisms. The time required for low and intermediate-mass stars ($M<8~\msun$) to reach the AGB phase ($10^8$~yr to few $10^9$~yr) is comparable to the age of the Universe at $z\simeq 6$ \citep{Morgan:2003}. The progenitors of core-collapse type II supernovae (SNII) are instead much more short-lived ($10^6$~yr) and therefore they must be the dominant dust sources at high-$z$ \citep[but see also][]{Valiante:2009}, provided that a significant fraction of grains survives the reverse shock \citep{Bianchi:2007, Hirashita2008}. This argument makes SNe the most attractive solution to explain the large amounts of dust $(M_{\rm dust} = 10^7-10^8~\msun)$ in high-$z$ quasars measured from their rest-frame far-infrared (FIR) emission when the Universe was only 1 Gyr old \citep[e.g.][]{bertoldi:2003dust, Michalowski2010, CarilliWalter2013, Gallerani:2017}. 

Evidence for SN-dust at high-redshift has been also reported in studies of gamma-ray bursts (GRBs) afterglows, as in the case of the GRB071025 afterglow at $z\sim 5$ \citep{Perley2010MNRAS, Jang2011ApJ} and GRB050904 at $z=6.3$ \citep{Stratta2007}. However, both these results are controversial \citep[e.g.][]{LiLiang2008, Zafar2010, Stratta:2011} and studies on larger samples of GRBs up to $z\sim 7$ showed either no significant extinction, or a variety of extinction curves compatible with the SMC, SMC-bar, LMC and MW \citep{Zafar2011, Zafar2018MNRAS, Zafar2018MNRASx, Bolmer2018}. Comparing these results with the ones found in AGN is not straightforward, because AGN-host galaxies might have a peculiar dust evolution history with respect to normal star-forming galaxies \citep[e.g.][]{Nozawa:2015}, or quasar spectra might reveal only the dust component in the AGN proximity, whereas GRBs probe the ISM of the host galaxy. Therefore, in order to understand the properties of dust in high-redshift quasars, dedicated theoretical models and observations are required.

In the last decades, many theoretical works investigated the evolution of chemical species and dust abundance in galaxies \citep[e.g.][]{Dwek1998ApJ, Zhukovska2008, Pipino2011}, but with the simplifying assumption of a single representative grain size. Recently, \citet{Asano:2013MNRAS} developed a framework fully including the grain size distribution evolution. They conclude that the latter quantity is fundamental to properly follow the dust mass evolution, as the efficiencies of all the physical mechanisms in the dust production/destruction cycle depend on the grain size. They also show that the grain size distribution evolves significantly throughout the galaxy evolution. Later on, similar models have been developed \citep[e.g.][]{Hirashita:2015, Nozawa:2015, Hirashita:2019}, and they have been used to make predictions for extinction curves of high-redshift galaxies. However, despite these efforts, many uncertainties in the theoretical models remain and a complete picture of the dust properties at high redshift is still missing. 

A complementary approach to the problem is to study the dust properties at high-redshift by post-processing hydrodynamical simulations with radiative transfer calculations \citep[e.g.][]{Hou:2017, Behrens:2018, Narayanan2018att, Aoyama:2020, ShenX2020}. This strategy has the advantage to combine the knowledge of the dust composition and grain size distribution of a theoretical model with the detailed gas/stars  distributions predicted by the hydrodynamical simulations. This allows to reliably compute the SED and the attenuation curve, which can then be compared with data, gaining insight about the dust content and composition of the observed source. However, most of these studies focused on normal star-forming galaxies, and the AGN contribution has often been neglected in this context. An exception is the work by \citet{Li2008ApJ}, who applied radiative transfer calculations to cosmological hydrodynamical simulation, taking into account also AGN radiation and different dust models. They were able to reproduce the SED of the quasar SDSS J1148+5251, and inferring its dust properties, finding that dust from supernova origin best explain the observations. However, this work was limited to the analysis of a single quasar.

In \citet{DiMascia:2021} (hereafter \citetalias{DiMascia:2021}), we studied the AGN contribution to the IR emission in AGN-host galaxies at high-redshift, by post-processing cosmological hydrodynamical simulations of $z\simeq 6$ quasars \citep[][hereafter {\citetalias{paramita:2018}}]{paramita:2018} with the radiative transfer code \code{SKIRT} \citep{Baes:2003, Baes:2015, Camps:2015, Camps2016}. In this work, we make use of the same simulations to investigate dust attenuation properties at high-redshift by comparing our synthetic SEDs with a large sample of bright $z\simeq 6$ quasars.

The paper is organised as follows: in Section \ref{Numerical_methods} we describe both the hydrodynamical simulations (Section \ref{hydro_sim}) and the model adopted for the radiative transfer calculations (Section \ref{RT_sim}). In Section \ref{sec:ext_curves} we present the attenuation curves obtained with our calculations, and investigate how their shape is affected by the AGN activity. In Section \ref{sec:obs_comparison} we compare our results with observations of $z\simeq 6$ quasars; in Section \ref{sec:best_model} we identify the best-fit dust model. In Section \ref{sec:discussion} we discuss the implications of our results and some caveats of the model adopted. Summary and conclusions are given in Section \ref{sec:conclusions}.

\section{Numerical model} \label{Numerical_methods}
The numerical model adopted in this work is similar to the one implemented in \citetalias{DiMascia:2021}. Below we summarise the common features and highlight differences between the \citetalias{DiMascia:2021} model and the one used here, both in terms of the hydrodynamical simulations (Section \ref{hydro_sim}), and the radiative transfer setup (Section \ref{RT_sim}).

\subsection{Hydrodynamical simulations} \label{hydro_sim}

We consider the hydrodynamical cosmological simulations studied in \citetalias{paramita:2018}, in which the evolution of a $\sim 10^{12}~\msun$ DM halo is followed from ${z=100}$ down to ${z=6}$ inside a comoving volume of ${(500~{\rm Mpc})^3}$ in a zoom-in fashion. The simulations are performed with a modified version of the Smooth Particle Hydrodynamics (SPH) N-body code \code{GADGET-3} \citep{Springel:2005}, accounting for stellar winds, supernovae feedback and metal enrichment. Radiative heating and cooling is included by using the tables computed in \citet{Wiersma2009MNRAS}. \citetalias{paramita:2018} adopts the multiphase model by \citet{Springel:2003}, which follows the hot and cold phase, resulting in an effective equation of state. Star formation is implemented with a density-based criterion, with a density threshold for star formation of ${n_{\rm SF} = 0.13 \ \cc}$ and assuming a \citet{Chabrier:2003} initial mass function (IMF) in the mass range ${0.1-100~\msun}$.
The mass resolution in the zoom-in region is ${m_{\rm DM} = 7.54 \times 10^6~\msun}$ and ${m_{\rm gas} = 1.41 \times 10^6~\msun}$ for DM and gas particles, respectively. The softening length for gravitational forces for the high-resolution DM and gas particles is $\epsilon = {1~ h^{-1}~{\rm kpc}}$ comoving.

BH evolution is also implemented, by seeding a ${M_{\rm BH} = 10^5~\msun}$ BH at the centre of a BH-less DM halo when it reaches a total mass of ${M_{\rm h} = 10^9~\msun}$. BH growth proceeds via gas accretion and mergers. The former is modelled via the classical Bondi-Hoyle-Littleton accretion rate ${\dot{M}_{\rm Bondi}}$ \citep{Hoyle:1939, Bondi:1944, Bondi:1952} and it is capped at the Eddington rate ${\dot{M}_{\rm Edd}}$. A fraction of the accreted rest-mass energy is radiated away with a bolometric luminosity
\begin{equation}\label{eq:luminosity_bh}
    L_{\rm bol} = \epsilon_{\rm r} \dot{M}_{\rm BH} c^2,
\end{equation}
where $c$ is the speed of light and $\epsilon_{\rm r}=0.1$ is the radiative efficiency. A fraction ${\epsilon_{\rm f} = 0.05}$ of the energy irradiated by the BH is distributed to the surrounding gas in kinetic form. 

In this work, we consider two runs performed in \citetalias{paramita:2018}: a run with kinetic feedback distributed inside a bi-cone with an half-opening angle of ${45\degree}$ (named \AGNcone{}), and a control simulation with no BHs (\noAGN{}). We focus on the ${z=6.3}$ snapshots of these runs, already analysed in \citetalias{DiMascia:2021}. The different physical properties among these runs in terms of gas morphology, star formation rate and black hole activity allow us to study the relative impact of dust and radiation sources distribution on the observed attenuation curves in AGN-host galaxies. For our analysis we select a cubic region of $60$ physical kpc size centred on the halo's centre of mass (the virial radius of the most massive halo is ${\approx 60}$~kpc at this redshift). We report the main physical properties of the zoomed-in halo inside this region in Table \ref{tab:hydro_runs}.

\begin{table*}
	\centering
	\begin{tabular*}{0.75\textwidth}{ c c c c c c c}
	\hline\noalign{\smallskip}
         run      & AGN feedback   & $M_{\rm gas}$ [$\msun$]   &  $M_\star$ [$\msun$]      & $\SFR$ [$\msunyr$]  & $\dot{M}_{\rm BH}$  [$\msunyr$] \\
         \hline\hline\noalign{\smallskip}
         \noAGN{}         & no         &   $2.9 \times 10^{11}$    &  $1.2 \times 10^{11}$ & $ 600$              &   -   &  \\
         \AGNcone{}       & bi-conical    &   $1.4 \times 10^{11}$    &  $7.0 \times 10^{10}$ & $ 189$              &   $89$ \\
	\hline
	\end{tabular*}
  	\caption{Summary of the main physical properties of the zoomed-in halo at ${z=6.3}$ within a cubic region of $60$~kpc size for the two hydrodynamic runs from \citetalias{paramita:2018} used in this work. For each run, we indicate: the feedback model used in the simulation, the gas mass ($M_{\rm gas}$), the stellar mass ($M_\star$), the star formation rate ($\SFR$, averaged over the last $10$~Myr), and the sum of the accretion rate of all the BHs in the selected region ($\dot{M}_{\rm BH}$).
  	\label{tab:hydro_runs}
  	}
\end{table*}

\subsection{Radiative transfer}\label{RT_sim}

We post-process the selected snapshots of the hydro simulations by using the publicly available code \code{SKIRT}\footnote{Version 8, \url{http://www.skirt.ugent.be}.} \citep{Baes:2003, Baes:2015, Camps:2015, Camps2016}. \code{SKIRT} is a Monte-Carlo radiative transfer solver, optimized to simulate radiation fields in dusty media, accounting for dust grains scattering and absorption, and their subsequent re-emission in the IR. Below we describe our implementation of dust distribution and properties (Section \ref{dust_model}), and radiation sources (Section \ref{radiation_sources}).

\subsubsection{Dust model} \label{dust_model}

Dust formation and evolution is not tracked in the hydrodynamic simulations considered here. We derive the dust mass distribution by assuming a linear scaling with the gas metallicity\footnote{Throughout this paper the gas metallicity is expressed in solar units, using ${\zsun=0.013}$ as a reference value \citep{Asplund:2009}.} \citep{Draine:2007}, parametrizing the mass fraction of metals locked into dust as:
\begin{equation} \label{eq:fd}
    f_{\rm d} = M_{\rm d} / M_Z,
\end{equation}
where $M_{\rm d}$ is the dust mass and $M_Z$ is the total mass of all the metals in each gas particle in the hydrodynamical simulation. We assume that gas particles hotter than $10^6$~K are dust-free because of thermal sputtering \citep{Draine:1979, Tielens:1994}.
The choice of $f_{\rm d}$ acts as a normalization factor for the total dust content. Its value is poorly constrained at high-redshift, both from observations and theoretical works (e.g. \citealt{Nozawa:2015,Wiseman:2017}). In this work we adopt a value (${f_d=0.08}$) tuned for hydro-simulations \citep{pallottini:2017althaea, Behrens:2018} to reproduce the observed SED of a ${z\sim 8}$ galaxy \citep{laporte:2017apj} and a MW-like value ($f_{\rm d} = 0.3$). 

\begin{figure*}
    \centering
    \hfill
    \includegraphics[width=0.975\textwidth]{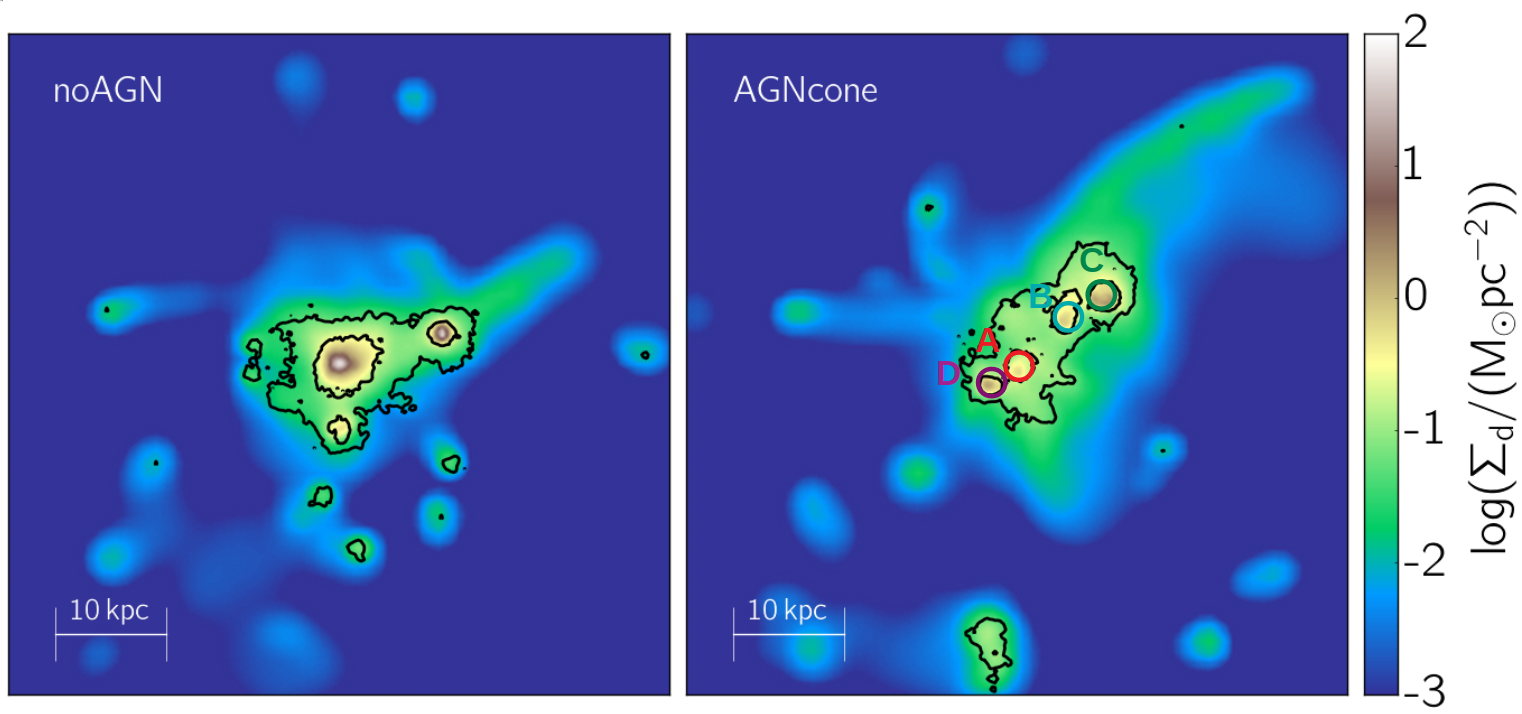}
    \hfill
    \caption{Dust surface density distribution of the most-massive halo at $z=6.3$ in \noAGN{} (left) and \AGNcone{} (right) for a cubic region with size $60$~kpc. A dust to metal ratio of $f_{\rm d}=0.08$ is assumed. Coloured circles indicates the position of three active BHs (source A, B, C) and a star-forming galaxy (source D), part of a merging system (see Section \ref{radiation_sources}). The black contours enclose regions with intrinsic UV emission higher than $S_{\rm UV} = 10^{9}~\lsun~{\rm kpc}^{-2}$ (outer contours) and $S_{\rm UV} = 10^{10}~\lsun~{\rm kpc}^{-2}$ (inner contours). 
    \label{fig:dust_map}
    }
\end{figure*}

Dust is distributed in the computational domain in an octree grid with a maximum of 8 levels of refinement for high dust density regions, achieving a spatial resolution of ${\approx 230}$~pc in the most refined cells, comparable with the softening length in the hydrodynamic simulation (${\approx 200}$ physical pc at $z=6.3$). In Figure \ref{fig:dust_map}, we show the dust distribution derived under these assumptions for our reference line of sight (los), namely the one aligned with the angular momentum of the particles inside the selected region. Regions with higher dust densities correspond to active star-forming regions, where gas metal enrichment is more effective. The dust distribution is affected by the AGN feedback, as also discussed in \citetalias{DiMascia:2021}. In \noAGN{}, the higher star formation rate and the absence of AGN feedback lead to a more compact dust distribution, with a surface density $\Sigma_{\rm d}$ that peaks at $\Sigma_{\rm d} \approx 50~\surfdpc$, whereas in \AGNcone{} it reaches $\Sigma_{\rm d}\approx 2~\surfdpc$. 

The dust chemical composition and its detailed grain size distribution in early (AGN-host) galaxies are still a matter of debate.  
We consider dust compositions and grain size distributions appropriate for the SMC and Milky Way (MW), by using the results of \citet{Weingartner:2001}. However, we consider also dust models derived from the \citet{Weingartner:2001} models by modifying the grain size distribution (see Section \ref{sec:obs_comparison}). We defer the inclusion of a SN-type extinction curve to a future work. We note that these extinction curves are somewhat intermediate between the SMC and MW curve, with the notable difference that SN-type curves are essentially flat in the wavelength range $1700-2500~\angstrom$. We sample in \code{SKIRT} the grain size distribution of graphite, silicates and polycyclic aromatic hydrocarbons (PAHs) using 5 bins for each component. In the SMC model the fraction of dust in PAHs is set to zero. 

Note that the assumed dust properties do not necessarily imply that the resulting \emph{attenuation} curves in our simulations match the extinction curve of the dust model used, because attenuation curves depend not only on the dust properties but also on the geometry of the distributions of dust and radiative sources and on radiative transfer effects \citep[e.g.][]{Witt96, Bianchi00, Behrens:2018, Narayanan2018att}. We discuss this point in detail in Section \ref{sec:ext_curves}.

We take into account dust self-absorption in our calculations. We instead neglect non-local thermal equilibrium (NLTE) corrections to dust emission. We do not include heating from CMB radiation. As discussed in \citetalias{DiMascia:2021}, only a small dust mass fraction in our simulations is at a temperature $\simeq T_{\rm CMB}$; therefore the corresponding corrections are negligible.

\subsubsection{Radiation sources} \label{radiation_sources}

Stars and black holes provide the UV radiation mainly responsible for dust heating. They are considered as point sources in our calculations, located at the position of the corresponding particle. The SED describing a stellar particle emission is derived according to the family of stellar synthesis models by \citet{Bruzual:2003}. For the black holes, we adopt the two AGN SED we implemented in \code{SKIRT} and described in \citetalias{DiMascia:2021}, based on a number of observational and theoretical works (\citealt{Shakura:1973, Fiore:1994, Richards2003AJ, Sazonov:2004, Piconcelli:2005}; \citetalias{Gallerani:2010}; \citealt{Lusso:2015, Shen:2020}). The AGN SED can be written as a composite power-law:
\begin{equation}\label{AGN_SED_eq2}
     L_\lambda = c_i \ \left(\frac{\lambda}{\mu{\rm m}}\right)^{\alpha_i} \ \left(\frac{L_{\rm bol}}{\lsun}\right) \ \lsun \ {\mum}^{-1},
\end{equation}
where $i$ labels the bands in which we decompose the spectra and the coefficients $c_i$ are determined by imposing the continuity of the function based on the slopes $\alpha_i$ (see Table 2 in \citetalias{DiMascia:2021}). In particular, we consider two AGN SEDs, which differ for the UV spectral slope, which is $\alpha_{\rm UV} = -1.5$ for the \emph{fiducial} case, and $\alpha_{\rm UV} = -2.3$ for the so-called \emph{UV-steep} case.
The SED is then normalized according to the bolometric luminosity of the BH, as expressed by eq. \ref{eq:luminosity_bh}.

The radiation field is sampled by using a grid covering the \emph{rest-frame} wavelength range ${[0.1-10^3]~\mum}$. The base wavelength grid is composed of 200 logarithmically spaced bins. If a MW-type dust is used, we add a nested grid with 200 logarithmically spaced bins in the wavelength range $[1-40]$ $\mum$ in order to better capture PAH emission (the composite wavelength grid has a total of 320 bins in this case). A total of $10^6$ photon packets\footnote{We verified that the number of photon packets used in our calculations is sufficient to reach convergence. We have compared the results with the ones from a simulations with ${5\times10^5}$ photons, and found negligible differences.} per wavelength bin is launched from each source. We collect the radiation escaping our computational domain for the six lines-of-sight perpendicular to the faces of the cubic computational domain. 

In Fig. \ref{fig:dust_map} we also mark with coloured circles the positions of the most accreting BHs in \AGNcone{}. These sources are labelled with the letters A (red circle, $\dot{M}_{\rm BH} \approx 32~\msunyr$), B (cyan circle, $\dot{M}_{\rm BH} \approx 7~\msunyr$), and C (green circle, $\dot{M}_{\rm BH} \approx 50~\msunyr$). We also mark the position of a star-forming galaxy (source D, purple circle). This merging system was studied in detail in \citetalias{DiMascia:2021} (see their Section 4.1). 

\section{Synthetic attenuation curves} \label{sec:ext_curves}

The extinction curve entangles information on the dust composition and grain size distribution, which determine how dust grains absorb and scatter photons at different wavelengths. The absolute value of the attenuation at a given wavelength depends also on the overall dust content. Moreover, the spatial distribution of radiative sources and absorbers play a crucial role in determining the observed flux via radiative transfer effects. The same extinction curve, i.e. the same dust properties, can produce different attenuation curves depending on the dust-sources geometry \citep[see e.g.][]{Witt96, Bianchi00, Seon2016ApJ, Behrens:2018, Narayanan2018att, Salim2020, Liang:IRXbeta}. Dust attenuation is therefore a complex process that depends on dust content, dust composition and on the geometry of the system.

\begin{figure*}
    \centering
    \vskip\baselineskip
    \includegraphics[width=0.475\textwidth]{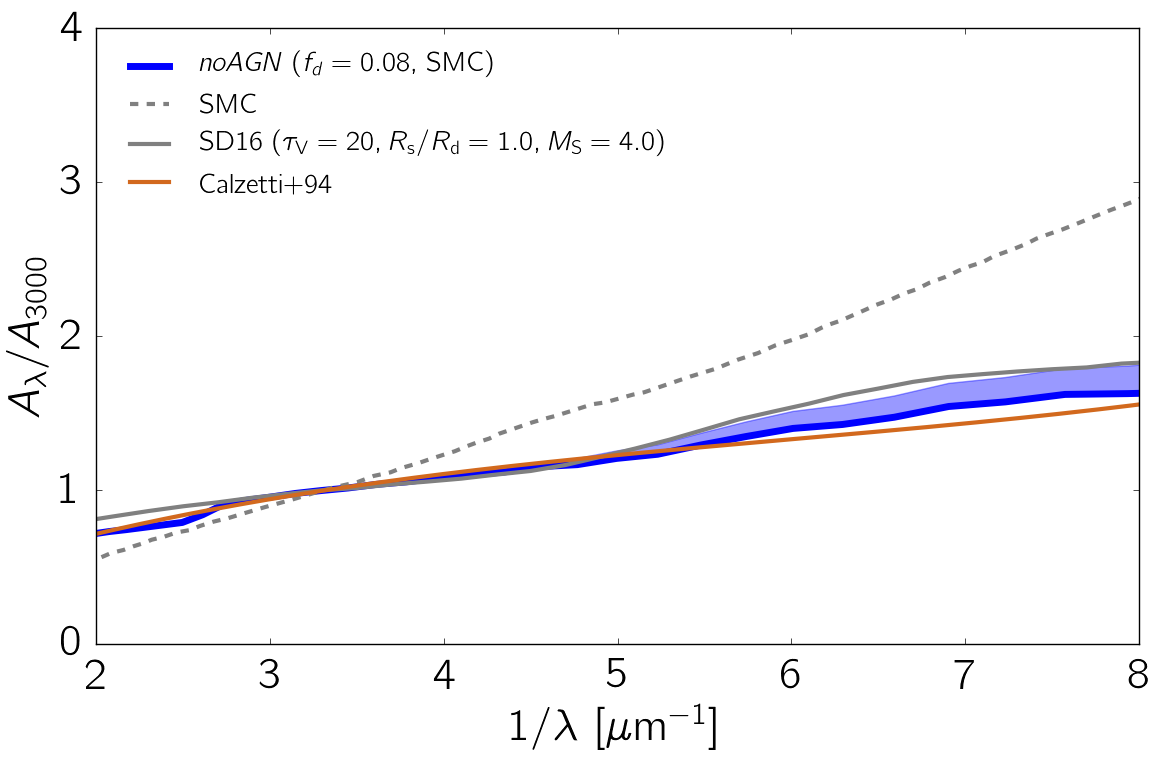}
    \hfill
    \includegraphics[width=0.475\textwidth]{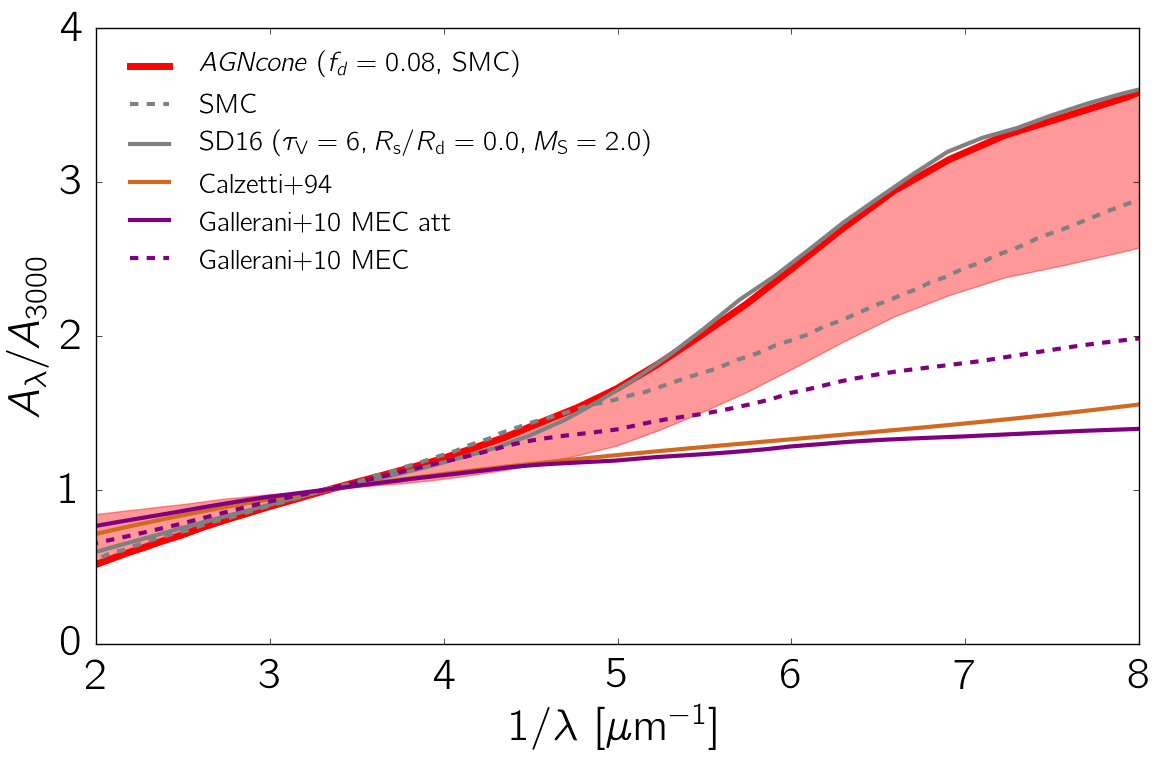}
    \vskip\baselineskip
    \includegraphics[width=0.475\textwidth]{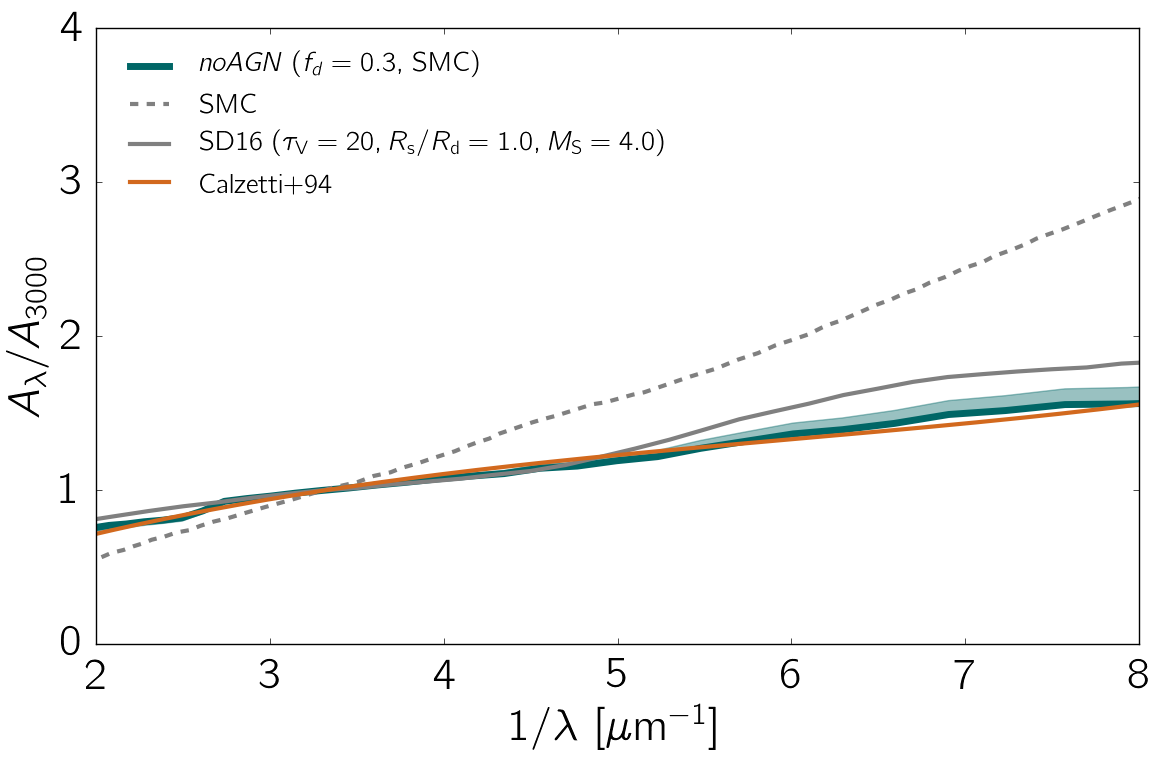}
    \hfill
    \includegraphics[width=0.475\textwidth]{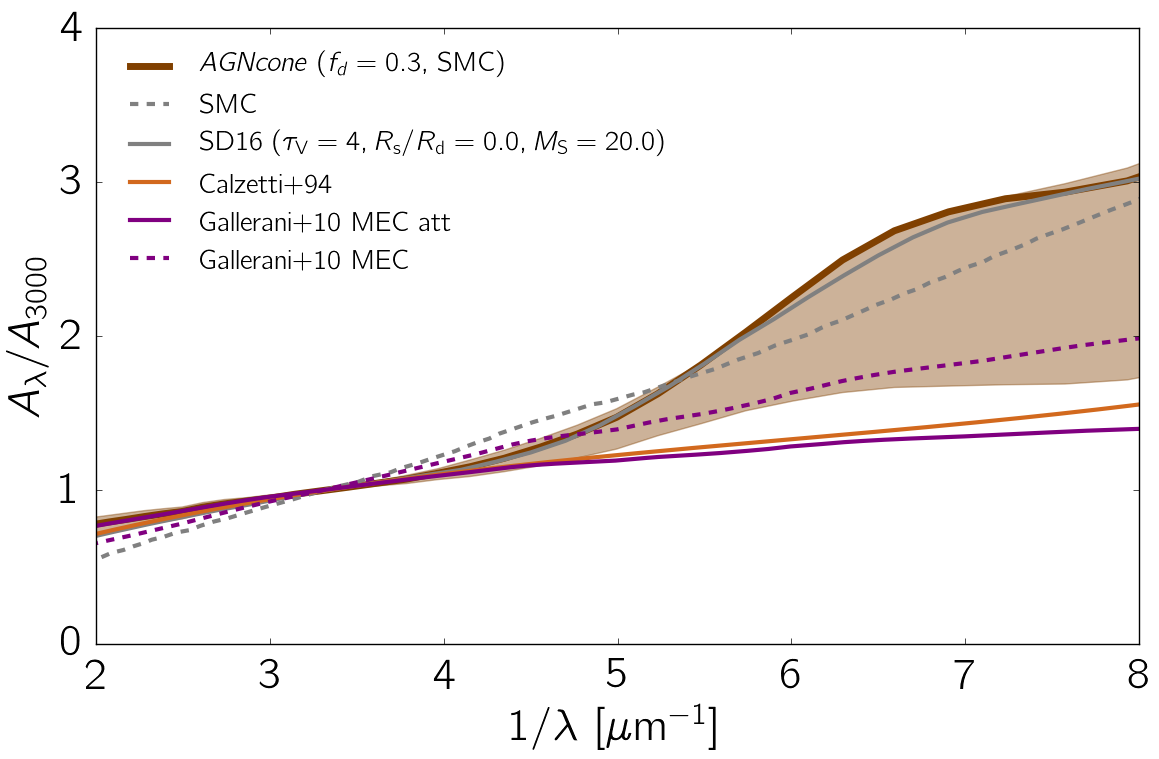}
    \vskip\baselineskip
    \caption{Attenuation curves (normalised to their value at $3000 \angstrom$) corresponding to four of the runs performed in \citetalias{DiMascia:2021} (see their Figure 8). The first column refer to \noAGN{}, the second to \AGNcone{}. The first row shows the case $f_{\rm d}=0.08$, the second row $f_{\rm d}=0.3$. The shaded area indicate the scatter between different lines of sight, used for the RT calculation. The solid line refers to our reference los. We also plot the SMC extinction curve (grey dashed line) and the Calzetti attenuation curve \citep[light brown solid line,][]{Calzetti:1994}. We also show with a grey solid line the attenuation curves derived in \citet{Seon2016ApJ} (assuming an SMC extinction curve) that best match our attenuation curves for the reference los, and we quote the corresponding parameters $(\tauV, R_{\rm s}/R_{\rm d}, M_{\rm S})$ describing the geometry adopted for that model (see text for more details). For \AGNcone{} we also show in purple the extinction curve (MEC, dashed line) and attenuation curve (MEC attenuated, solid line) obtained in \citetalias{Gallerani:2010} for high-redshift quasars.
    \label{fig:attenuation_curves} 
    }
\end{figure*}

Fig. \ref{fig:attenuation_curves} shows the attenuation curves resulting from the RT simulations performed by \citetalias{DiMascia:2021} for the runs \emph{noAGN${008}$}, \emph{noAGN${03}$}, \emph{AGNcone$008$}, \emph{AGNcone$03$}. The \noAGN{} run (left panels) is representative of a high star-forming ($\SFR \approx 600~\msunyr$) $z\sim 6$ galaxy, whereas the \AGNcone{} run (right panels) is representative of a bright, $M_{UV}\sim -26$, $z\sim 6$ quasar. This comparison allows us to study the role of AGN (if any) in shaping the attenuation curve of its host galaxy. The attenuation curves shown are obtained by normalizing $A_{\lambda}$ to its value at 3000 $\angstrom$, where $A_{\lambda}=1.086\tau_{\lambda}$, and $\tau_{\lambda}=-\ln(F_{\lambda}^{\rm obs}/F_{\lambda}^{\rm int})$, with $F_{\lambda}^{\rm obs}$ and $F_{\lambda}^{\rm int}$ being the observed and the intrinsic flux, respectively.
In these simulations an intrinsic SMC-like extinction curve is adopted. In the next two subsections, we investigate the origin of the different attenuation curves, by focusing on the slope and the los-to-los variations.

\subsection{Slope of the attenuation curve} \label{sec:slope_att}

The attenuation curves in the \noAGN{} run are flatter than in \AGNcone{}, for a fixed dust-to-metal ratio. In particular, the ratio of the attenuation at 0.1 ($A_{0.1}$) and 1 $\mum$  ($A_1$) rest-frame is  $A_{0.1}/A_1\approx 3-4$ in \noAGN{}, whereas it can be as high as $\approx 10$ in \AGNcone{} runs. The flattening of the attenuation curve in the \noAGN{} case is due to the high dust optical depth characterising this run ($\Sigma_{\rm d} \approx 50~\surfdpc$ for $f_{\rm d}=0.08$, see Fig. \ref{fig:dust_map}). Regions characterized by these large surface densities become optically thick even to rest-frame NIR photons. This is evident by comparing the intrinsic and observed flux in the SED (see Fig. 8 in \citetalias{DiMascia:2021}). Instead, the dust surface density is at most $\approx 2~\surfdpc$ in \AGNcone{}, such that the attenuation is significant at the shortest wavelengths (where the extinction is higher because of the dust optical properties), but the dust is optically thin in the NIR.

The importance of the dust surface density in shaping the attenuation curve in our simulations can also be appreciated by comparing the results of the same run for different dust contents. In fact, for the reference los, the ratio between the attenuation at $0.1~\mum$, $A_{0.1}$, and the attenuation at $1~\mum$, $A_1$,  decreases in \AGNcone{} from the case $f_{\rm d}=0.08$ to $f_{\rm d}=0.3$. As a result, the attenuation curves in \AGNcone{} tend to become flatter when increasing the dust content.

We also compare our predicted attenuation curves with the \citet{Calzetti:1994} curve, and we find that \noAGN{} is in very good agreement, whereas the curves in \AGNcone{} are much steeper. We also consider the results obtained by \citet{Seon2016ApJ}, which performed radiative transfer simulations of a single turbulent cloud, in order to study how the geometry of the cloud shapes the attenuation curves at fixed extinction curve (i.e. SMC, MW, LMC). A comparison between our results and simulations of a single cloud is quite difficult, but it can give useful insights. In the \citet{Seon2016ApJ} models, dust and stars are assumed to be distributed in spherical symmetry inside a turbulent cloud (characterised by a Mach number $M_{\rm S}$), within a radius $R_{\rm d}$ and $R_{\rm s}$ respectively. $R_{\rm s}=0$ represents a cloud with a bright point source at its center, whereas $R_{\rm s}/R_{\rm d}=1$ corresponds to a case in which dust and stars are uniformly distributed. A model is uniquely determined by the triplet ($\tauV$, $R_{\rm s}/R_{\rm d}$, $M_{\rm S}$), where $\tauV$ is the V-band optical depth.

In each panel of Fig. \ref{fig:attenuation_curves} we show the model in \citet{Seon2016ApJ} that best fits the attenuation curve for the reference los, based on a minimum squares criterion. For the \noAGN{} run, we find that this model is characterised by a high optical depth ($\tauV=20$, which is the maximum value adopted in their work) and by $R_{\rm s}/R_{\rm d}=1$. This further confirms our explanation that the flatness of the attenuation curves in \noAGN{} is determined by a very high dust surface density. The ratio $R_{\rm s}/R_{\rm d}=1$, corresponding to the case of stars and dust well mixed, is consistent with the compactness of the galaxies in \noAGN{} (see also the left panel in Fig. \ref{fig:dust_map}). 
For the \AGNcone{} run, the best-fit model by \citet{Seon2016ApJ} has a lower optical depth $\tauV=4$ with respect to \noAGN{}, which is again consistent with the dust surface density, and a ratio $R_{\rm s}/R_{\rm d}=0$. This case corresponds to a geometry with a bright point source at the center of the cloud, which is a reasonable approximation for the geometry in the \AGNcone{}.

The relation between the dust optical depth and the steepness of the attenuation curves, and how they change among the different simulations, can be understood with the following analytical argument. For this calculation, we make use of two simplified geometries, but it is useful to get a physical insight about how the dust optical depth affects the shape of the attenuation curves.

The attenuation at a specific wavelength $\lambda$ is defined as: $A_\lambda = 1.086 \tau_\lambda = - 1.086 \ln(T (\tau_\lambda, \zeta_\lambda, \xi_\lambda))$, where $T$ is the \emph{transmissivity}, i.e. the fraction of emergent light. $\zeta_\lambda$ and $\xi_\lambda$ are functions of the albedo $\omega(\lambda)$ and the asymmetry parameter $g(\lambda)$ of the Henyey-Greenstein scattering phase function at the wavelength considered. The albedo is defined as $\omega(\lambda)=\sigma_{\rm scat}(\lambda)/\sigma_{\rm ext}(\lambda)$, where $\sigma_{\rm scat}(\lambda)$ and $\sigma_{\rm ext}(\lambda)$ are the scattering and extinction cross sections at the wavelength $\lambda$, respectively. 
We consider two different geometries: i) a point source surrounded by dust in a spherically symmetric distribution (\emph{point source model}); ii) a sphere with dust and emitters uniformly distributed (\emph{intermixed model}). For the first case, we adopt the classical solution for the transmissivity derived by \citet{Code1973IAUS}:
\begin{equation} \label{eq:transmissivity_Code}
    T_{\rm ps} (\tau_\lambda, \zeta_\lambda, \xi_\lambda) = \frac{2}{(1+\zeta_\lambda)e^{\xi_\lambda \tau_\lambda} + (1-\zeta_\lambda)e^{-\xi_\lambda \tau_\lambda}},
\end{equation}
where the functions $\zeta_\lambda$ and $\xi_\lambda$ have the following functional form:
\begin{align}
    &\zeta_\lambda(\omega_\lambda, g_\lambda) = \sqrt{(1-\omega_\lambda)/(1-\omega_\lambda g_\lambda)} \\
    &\xi_\lambda(\omega_\lambda, g_\lambda) = \sqrt{(1-\omega_\lambda)(1-\omega_\lambda g_\lambda)}.
\end{align}
For the homogenous sphere, we make use of the solution obtained by \citet{Osterbrock1989} (see also Appendix C in \citealt{Varosi1999ApJ}):
\begin{equation} \label{eq:transmissivity_Osterbrock}
    T_{\rm mix} (\tau_\lambda) =  \frac{3}{4\tau_\lambda} \left\{1-\frac{1}{2\tau^2_\lambda} + \left(\frac{1}{\tau_\lambda} + \frac{1}{2\tau^2_\lambda} \right)e^{-2\tau_\lambda}\right\},
\end{equation}
which expresses the escaping probability for photons in a sphere where sources and absorbers are homogeneously distributed and the opacity from the centre to the surface is $\tau_\lambda$.
We then rewrite eq. \ref{eq:transmissivity_Code} and \ref{eq:transmissivity_Osterbrock} by expressing the optical depth $\tau_\lambda$ in terms of the optical depth in the V-band, $\tau_\lambda = (\sigma_{{\rm ext}},\lambda/\sigma_{\rm ext, V}) \tauV$, and making use of the formulas above, we can write the attenuation at a given wavelength $A_{\lambda}$ in terms of the V-band optical depth.

\begin{figure*}
	\centering
	\includegraphics[width=0.45\textwidth]{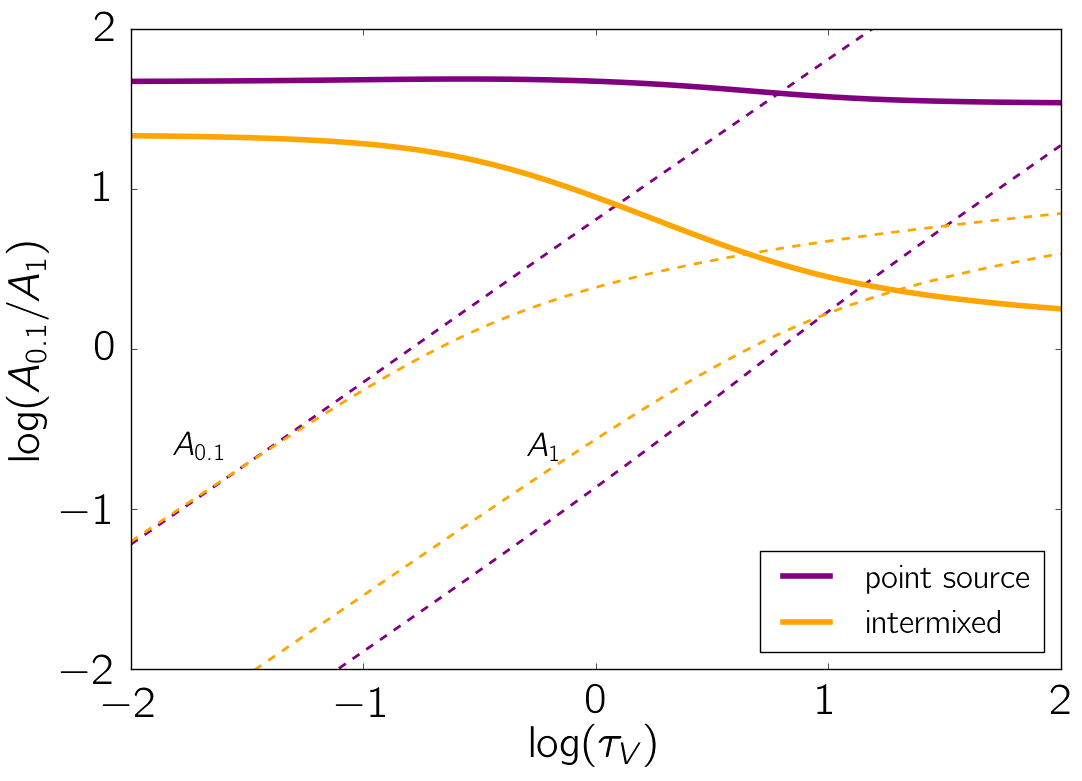}
	\hfill
	\includegraphics[width=0.452\textwidth]{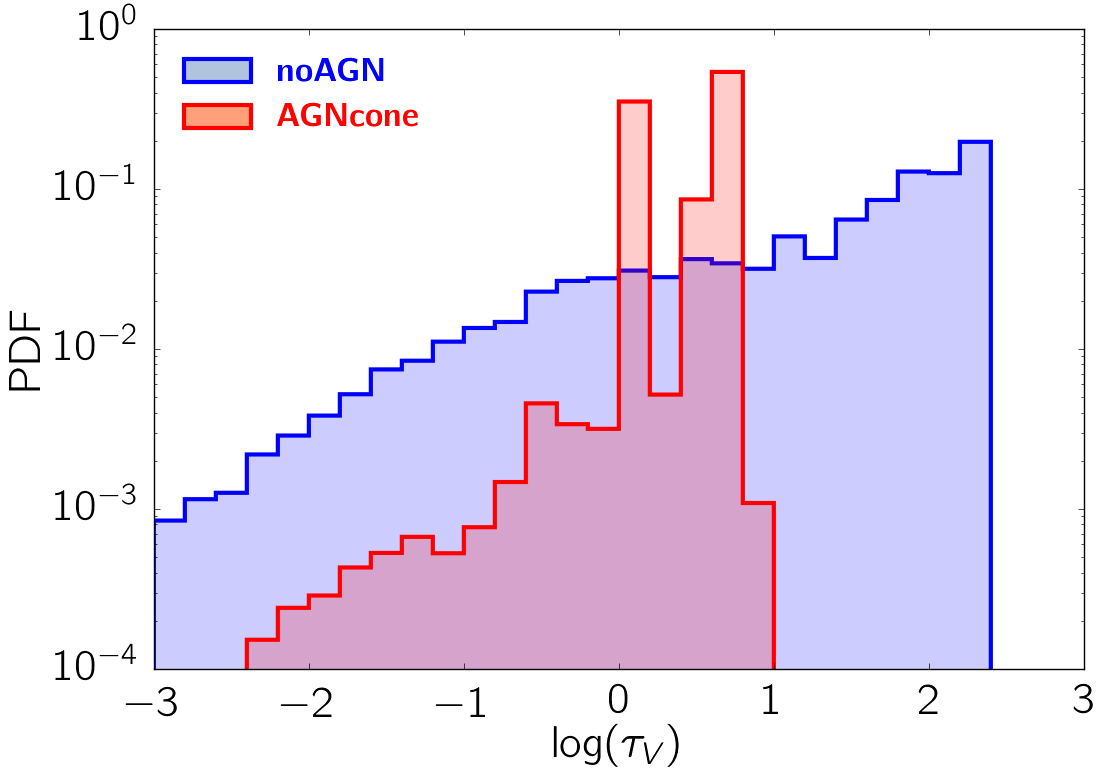}
	\caption{\emph{Left panel:} Ratio (solid lines) of the attenuations $A_{0.1}$ and $A_1$ (dashed lines), at $0.1~\mum$ and $1~\mum$ respectively, as a function of the V-band optical depth $\tauV$. The attenuation is computed from equations \ref{eq:transmissivity_Code} and \ref{eq:transmissivity_Osterbrock}, for the point source model (purple line) and the intermixed model (orange line) respectively, adopting parameter values appropriate for SMC dust. \emph{Right panel}: PDF of the \emph{slab} dust optical depth derived from the dust density distribution in the V-band, $\tauV$, for the reference los, assuming $f_{\rm d}=0.08$. The PDF is weighted by the intrinsic UV emissivity in the field of view. The histograms show the case \noAGN{} (blue) and \AGNcone{} (red). 
	\label{fig:attenuation_vs_optical_depth}
	}
\end{figure*}

In order to explore how the attenuation changes with optical depth, both in the optical/UV regime and in the NIR, we compute the attenuation at $0.1~\mum$ and $1~\mum$. For this calculation, we adopt the appropriate values for $\omega_\lambda$, $g_\lambda$ and $\sigma_{\rm ext}$ according to the results by \citet{Draine2003ApJ} for the SMC model (see their Fig. 4). In the left panel of Fig. \ref{fig:attenuation_vs_optical_depth} we plot the ratio $A_{0.1}/A_1$ as a function of $\tauV$ with thick solid lines, for the point source (purple) and intermixed (orange) model; dashed lines represent $A_{0.1}$ and $A_1$, separately.

At low optical depths, the ratio $A_{0.1}/A_1$ converges to a constant in both cases, which is a function only of the optical properties of the dust mixture assumed. In particular,
\begin{align*}
    &\frac{A_{0.1}}{A_1} \stackrel{\tauV \ll 1}{\longrightarrow} \frac{\zeta_{0.1} \xi_{0.1} \sigma_{0.1}}{\zeta_1 \xi_1 \sigma_1 } \simeq 46.87  \ \ \ \ &\mbox{(point source)} \\
    &\frac{A_{0.1}}{A_1} \stackrel{\tauV \ll 1}{\longrightarrow} \frac{\sigma_{0.1}}{\sigma_1} \simeq 21.81 \ \ \ &\mbox{(intermixed),}
\end{align*}
where $\zeta_\lambda$, $\xi_\lambda$, $\sigma_\lambda$ have been evaluated at the wavelength indicated in the subscript. At high optical depths, the ratio converges to: 
\begin{align*}
    &\frac{A_{0.1}}{A_1} \stackrel{\tauV \gg 1}{\longrightarrow} \frac{\sigma_{0.1}}{\sigma_1} \simeq 21.81 \ \ \ \ &\mbox{(point source)}  \\
    &\frac{A_{0.1}}{A_1} \stackrel{\tauV \gg 1}{\longrightarrow} 1 \ \ \ \ &\mbox{(intermixed).} 
\end{align*}

In both cases, the ratio $A_{0.1}/A_1$ decreases with increasing optical depth. This effect is limited in the point source model, whereas it is much more significant in the homogeneous sphere model, for which this ratio tends to $1$, corresponding to a flat attenuation curve.

We compare this analytical prediction with our simulations. We convert the dust surface density into an optical depth by using the extinction coefficient $\kappa_V = 1.72 \times 10^4~\extccg$. The maximum value of the V-band optical depth estimated in this way over the field of view for the chosen line of sight is $\approx 200$ and $\approx 7$ for \noAGN{} and \AGNcone{} respectively, assuming $f_{\rm d}=0.08$. 

We emphasize that the quoted opacity is the \emph{slab} opacity, i.e. it represents the attenuation expected for a light beam passing through a slab with the assumed dust density and properties. The corresponding attenuation would read simply as $e^{-\tau}$, with $\tau$ being the opacity considered. However the \emph{effective} opacity depends also on two essential ingredients: i) radiative transfer effects; ii) the relative distribution of dust grains and radiation sources. Both these factors can contribute in making the effective opacity much lower than the slab one.

Therefore, in order to make a more fair comparison of the results from the RT simulations (which include all the physical processes in play) with our analytical calculation (which include RT effects, but it is limited to a point-source geometry), we need to examine in more details how the slab opacity is distributed in the hydrodynamical runs. In fact, for a given line of sight, the optical depth varies from region to region, while the attenuation curves shown in Fig. \ref{fig:attenuation_curves} refer to the whole field of view. 

We compute the Probability Distribution Function (PDF) of the slab opacity from the dust distribution maps in Fig. \ref{fig:dust_map}. We perform the calculation for the reference line of sight, in order to directly compare the results with the estimates provided above. We restrict the computation to the subregion of the field of view for which the slab V-band optical depth is $\tauV \gtrsim 10^{-3}$, and we compute the PDF by weighting $\tauV$ with the \emph{intrinsic} UV emissivity, in order to emphasize the contribution from the regions responsible for the emission.

In the right panel of Fig. \ref{fig:attenuation_vs_optical_depth} we show the UV-weighted PDF of the slab $\tauV$ for \noAGN{} (blue) and \AGNcone{} (red), for the selected subregion of the field of view. The low density regions ($\tauV \ll 1$) contribute only to a small fraction of the weighted distribution, i.e. $\lesssim 15\%$ in \noAGN{} and $\lesssim 1\%$ in \AGNcone{}. The weighted fraction of dust in the range $1 < \tauV < 10$ is $\approx 16\%$ in \noAGN{}, whereas it constitutes almost the totality ($\approx 98\%$) in \AGNcone{}. This is because the dusty regions around active BHs in \AGNcone{}, which are the most UV-emitting regions (see black contours in Fig. \ref{fig:dust_map}), are characterised by optical depths in this range. In \noAGN{}, the most star-forming regions have $\tauV > 10$, which account for $\approx 69\%$ of the weighted distribution. Overall, the fraction of the field of view observed that has very high slab opacity is much higher in \noAGN{} than in \AGNcone{}. The low $A_{0.1}/A_1$ ratio in \noAGN{} suggests that the sphere model is a reasonable approximation of the simulated object, characterised by a compact distribution. This is also consistent with the comparison with the attenuation curves derived by \citet{Seon2016ApJ}. \AGNcone{} is instead better described by a combination of the point source and intermixed models. In fact, the point source model alone predicts a $A_{0.1}/A_1$ ratio a factor of $\approx 4$ higher than what we find from the simulations in the relevant optical depth range, $1 < \tauV < 10$. Despite the simplicity of these two analytical models, they are still able to explain why the resulting attenuation curve in $\noAGN{}$ is almost flat, whereas the ratio $A_{0.1}/A_1$ reaches up to $\approx 10$ in AGN runs.

The different optical depth distribution between the \noAGN{} and \AGNcone{} runs is the result of AGN feedback effects. The kinetic feedback prescription adopted in \citet{paramita:2018} (see Section \ref{hydro_sim}) generates powerful outflows that affect the gas distribution (and therefore the dust distribution) in the host galaxy, removing gas from the central regions and distributing it over several kpc, causing a more diffuse gas distribution as compared with the \noAGN{} case. Therefore, the steep attenuation curves we predict for AGN-host galaxies are a direct consequence of a gas distribution affected by AGN activity. We further discuss this point in Section \ref{sec:dust_spatial_disc}.

\subsection{Line of sight attenuation variations} \label{sec:disp_att}

The los-to-los variation of the slope of the attenuation curves is much smaller in \noAGN{} than in AGN runs, as can be seen from the width of the shaded regions in Fig. \ref{fig:attenuation_curves}. In fact, $A_{0.1}/A_1\approx 3-4$ in \noAGN{}, while it is $\approx 6-9$ in \AGNcone{}, for the $f_{\rm d}=0.08$ case. This scatter is also due to the role of AGN in shaping the gas distribution in its surrounding\footnote{We underline that we are focusing here on how the attenuation curve changes with the line of sight, whereas in Section \ref{sec:slope_att} we focus on the slope of the attenuation curve for a single los.}. In the \noAGN{} case, the gas distribution is more compact and uniform around star-forming regions with respect to \AGNcone{}. As a result, not only the attenuation curve for each los is flat, but also the scatter between each los is small. In \AGNcone{}, feedback from accreting BHs alters the surrounding matter distribution by cleaning gas along some directions: some los are characterised by low dust column densities, resulting into steep attenuation curves, as the extreme ones in \AGNcone{}, some other by high dust column densities, providing flat attenuation curve, similar to the ones found in \noAGN{}.

\section{Comparison with quasar data} \label{sec:obs_comparison}

In Section \ref{sec:ext_curves}, we studied the attenuation curves predicted by our simulations from a general point of view, in order to get insight about how the AGN activity affects the shape of the attenuation curves. In this Section we compare our results with observations of high-redshift quasars.

In Fig. \ref{fig:best models}, we show the synthetic SEDs obtained from our RT calculations for the \AGNcone{} run (${M_{\rm UV}=-27.97}$ before accounting for dust attenuation) and we compare them with observations of ${z\geq 6}$ bright (${-29\lesssim M_{\rm UV}\lesssim -26}$) quasars (see Table \ref{tab:sources_comparison} for all the sources of the sample) in the rest-frame UV-to-FIR (see also Fig. 9 in \citetalias{DiMascia:2021}). We find qualitatively an overall agreement between the results of our model with $f_{\rm d} = 0.3$ and the fiducial AGN SED\footnote{In \citetalias{DiMascia:2021} we considered two values for the dust-to-metal ratio, $f_{\rm d} = 0.08$ and $f_{\rm d} = 0.3$, an SMC-like dust composition and both the fiducial and the UV-steep AGN SED.} with the photometric data points, with a notable exception. In the optical/UV band, the predicted SEDs luminosities fall short with respect to the data by \citetalias{Gallerani:2010}; also, they show a steep decline in the far-UV, which is not observed. Different dust models might help in solving these mismatches. In fact, \citetalias{Gallerani:2010} fitted the spectra with extinction curves flatter than the SMC, hinting at different dust composition and/or grain size distribution at high-redshift.

\begin{table}
	\centering
	{\def\arraystretch{1.2}
	\begin{tabular}{ l c c c}
		\hline\noalign{\smallskip}
		source & $z$ & $M_{\rm UV}$ & Reference\\
		\hline\hline\noalign{\smallskip}
		J1030+0524 & 6.31& -27.12& [1-2,8]\\ 
		J1048+4637 & 6.23& -27.60& [1-2,8]\\ 
		J1148+5251 & 6.43& -27.85& [1-8]\\ 
		J1306+0356 & 6.03& -26.76& [1-2,8-9]\\ 
		J1602+4228 & 6.07& -26.85& [1-2,8]\\ 
		J1623+3112 & 6.25& -26.71& [1-2,8]\\ 
		J1630+4012 & 6.07& -26.16& [1-2,8]\\
		J0353+0104 & 6.07& -26.56& [8]\\ 
		J0818+1722 & 6.00& -27.44& [8]\\ 
		J0842+1218 & 6.08& -26.85& [8,9]\\ 
		J1137+3549 & 6.01& -27.15& [8]\\ 
		J1250+3130 & 6.13& -27.18& [8]\\ 
		J1427+3312 & 6.12& -26.48& [8]\\ 
		J2054-0005 & 6.04& -26.15& [8]\\ 
		P007+04 & 6.00& -26.58& [9]\\ 
		P009-10 & 6.00& -26.50& [9]\\ 
		J0142-3327 & 6.34& -27.76& [9]\\ 
		P065-26 & 6.19& -27.21& [9]\\ 
		P065-19 & 6.12& -26.57& [9]\\ 
		J0454-4448 & 6.06& -26.41& [9]\\ 
		P159-02 & 6.38& -26.74& [9]\\ 
		J1048-0109 & 6.68& -25.96& [9]\\ 
		J1148+0702 & 6.34& -26.43& [9]\\ 
		J1207+0630 & 6.04& -26.57& [9]\\ 
		P183+05 & 6.44& -26.99& [9]\\ 
		P217-16 & 6.15& -26.89& [9]\\ 
		J1509-1749 & 6.12& -27.09& [9]\\ 
		P231-20 & 6.59& -27.14& [9]\\ 
		P308-21 & 6.23& -26.30& [9]\\ 
		J2211-3206 & 6.34& -26.65& [9]\\ 
		J2318-3113 & 6.44& -26.06& [9]\\ 
		J2318-3029 & 6.15& -26.16& [9]\\ 
		P359-06 & 6.17& -26.74& [9]\\
		J0100+2802 & 6.33& -29.30& [10]\\
  	    P338+29 & 6.66& -26.01& [14]\\
  	    J0305-3150 & 6.61& -26.13& [15]\\
	\hline\end{tabular}
	}
  	\caption{Quasars used for the comparison with the prediction by our model. Columns indicate: (first) source name, (second) redshift, (third) $M_{\rm UV}$ and (fourth) references for the photometric data used for the $\chi^2$ analysis: [1] \citetalias{Gallerani:2010}; [2] \citet{Juarez:2009}; [3] \citet{Walter:2003}; [4] \citet{Bertoldi:2003}; [5] \citet{Riechers:2009}; [6] \citet{Gallerani:2014}; [7] \citet{Stefan:2015}; [8] \citet{Leipski2014ApJ}; [9] \citet{Venemans:2018}; [10] \citep{Wang:2016ApJ}; [11] \citep{Venemans:2012ApJ}; [12] \citep{Venemans:2017ApJ}; [13] \citep{Willott:2017ApJ}; [14] \citep{Mazzucchelli:2017ApJ}; [15] \citep{Venemans2016ApJ}. 
  	\label{tab:sources_comparison}
  	  	}
\end{table}

In order to verify this working hypothesis, we considered several models, which we summarised in the following.
We assume the dust-to-metal ratio to vary between $f_{\rm d} = 0.08$ and $f_{\rm d} = 0.3$, by considering also an intermediate value $f_{\rm d} = 0.15$, in addition to the ones considered in the previous analysis.
Given that a steep slope ($-2.9<\alpha_{UV}<-2.3$) is suggested in the analysis of reddened high-$z$ quasars by \citetalias{Gallerani:2010} (see their Table 2), we include both the fiducial and UV-steep AGN SEDs in this calculation (see Section \ref{radiation_sources}).
We adopt a grain size distribution and composition that match the extinction properties either of the SMC or the MW (the latter was not included in the previous work). We also consider alternative dust models, which have the same dust composition of the fiducial SMC/MW cases, but with a cut in the dust grain distribution at $\amin$. We use $\amin=0.01~\mum$ and $\amin=0.1~\mum$. We note that the grain size distribution of the SMC/MW dust models by \citet{Weingartner:2001} implemented in \code{SKIRT} extends down to $\amin=0.001~\mum$ for silicates and carbonaceous grains, and down to $\amin=0.0003548~\mum$ for PAH molecules. Grain size distributions deficient in small dust grains are suggested by flat attenuation curves measured in AGN \citep[e.g.][]{Gaskell:2004, Czerny:2004, Gallerani:2010} and justified by theoretical models \citep[e.g.][]{Hirashita:2019, Nozawa:2015}.

We end up with a total of 20 models, which are summarised in Table \ref{tab:chisquare_test}. We show as a reference the \emph{total} SED of these models in the left-hand panel in Fig. \ref{fig:best models}.

\subsection{Best-fit model} \label{sec:best_model}

In order to find the model that best matches the observational data, we perform a $\chi^2$ analysis by comparing our synthetic SEDs with the observational data of the quasars selected. In the following, we will refer to rest-frame wavelengths $1\lesssim \lambda \lesssim 5~\mum$ as Near-Infrared (NIR), $5\lesssim \lambda \lesssim 40~\mum$ as Mid-Infrared (MIR) and $40\lesssim \lambda \lesssim 350~\mum$ as Far-Infrared (FIR).

We first group the data in different bins: two bins for the rest-frame UV/optical wavelengths ($0.145~\mum$ and $0.3~\mum$ rest-frame), ten bins for the NIR/MIR/FIR corresponding to the photometric bands of the Spitzer/Herschel telescopes, and a single bin for the rest-frame FIR, at $1.18$~mm\footnote{The rest-frame optical/UV data for the quasars in the sample are the spectra studied in \citetalias{Gallerani:2010} taken with the TNG/GEMINI telescopes. We take the flux at rest-frame $0.145~\mum$ and $0.3~\mum$ from these spectra. Regarding the rest-frame FIR measures by ALMA, we collect into a single bin the points with observed wavelengths between $1.1$~mm and $1.3$~mm. Then, we simply assume that they are all taken at the average wavelength of the sample, namely $1.18$~mm. In the other cases the wavelength of the bin simply corresponds to the observed wavelength.}.
However, our models do not include the emission from hot dust in the torus around AGN, which might be significant in the NIR, as we also pointed out in \citetalias{DiMascia:2021} (see Figure 11). To overcome this limitation, we exclude from our analysis those bins characterised by a torus contamination above $5\%$\footnote{We model the torus emission as a single-temperature greybody, with dust mass $M_{\rm torus}=10~\msun$, $T_{\rm torus}=1000$~K and emissivity $\beta_{\rm torus}=2$. The contamination from the torus is then computed as the ratio between the flux from the torus component to the total (torus plus model) flux. We acknowledge that this is a simplification of the torus physical properties, whose temperature is expected to follow a power-law profile $\propto T_{\rm torus}^{-3/4}$, with a maximum corresponding to the sublimation temperature of the grains at $\approx 1500-1800$~K \citep[e.g.][]{Netzer:2015}. However, we find that the wavelength range mostly affected by the torus emission is consistent with what found in SED fitting of AGN including a torus component \citep[e.g.][]{Hernan-Caballero2016MNRAS}.} This procedure excludes the bins at $8~\mum$ and $24~\mum$. However, we argue that our results are not much affected by this choice, because the strongest constraints to the $\chi^2$ fitting come from the rest-frame UV/optical data, as discussed below.
We end up with a total of eleven bins at the following wavelengths: $0.145~\mum$ (rest-frame), $0.3~\mum$ (rest-frame), $3.6~\mum$, $4.5~\mum$, $5.8~\mum$, $100~\mum$, $160~\mum$, $250~\mum$, $350~\mum$, $500~\mum$, $1.18$~mm.
We assign to each bin an \emph{average flux estimate}, given by the mean value of all the fluxes in the bin, and an error, given by the standard deviation. Given that the fluxes span few order of magnitudes, we consider for each bin the logarithm of the flux. For what concerns the optical/UV bins, since errors in the observed spectra are not provided at these wavelengths, we assume a $10\%$ error on the average value of the logarithm of the flux. Furthermore, given that if a bin has few sources, the standard deviation might produce very small errors, we impose a minimum relative error for the average measure equal to $10\%$ of the mean value for that bin.

For what concerns the comparison between our predicted SEDs and optical/NIR/FIR observations, we first remind that our simulated field of view contains a merging system, composed by three AGN, which we labelled as A, B, C (see Section \ref{radiation_sources} and Fig. \ref{fig:dust_map}) and one star-forming galaxy (source D). We use the SED of source (A) at $0.145~\mum$ and $0.3~\mum$ rest-frame, and the SED of source (C) at $1.18$~mm\footnote{The SED of an individual source is extracted from a circle of $2.5$~kpc centred at the position of the source in the field of view.}, since they are the brightest in the UV and IR, respectively (for a detailed discussion, see Section 4.1 in \citetalias{DiMascia:2021}). For what concerns the NIR/MIR/FIR, we have to take into account that the resolution of Spitzer and Herschel data does not allow to resolve the two dominant emitting regions\footnote{Source (A) is in fact only $\sim 10$~kpc away from source (C).}. Thus, for the bins from $3.6~\mum$ to $500~\mum$, we compare the data with the total synthetic SED.

Finally, we compute the reduced $\chi^2$ between the binned data and the predictions by our models. In particular, we use the following formula: 
\begin{equation}
    \chi^2 = \frac{1}{N} \ \sum \left(\frac{\log F_{\rm model}-F^{\log}_{\rm obs}}{\sigma^{\log}_{\rm obs}}\right)^2,
\end{equation}
where the sum is performed over the total N bins, $F_{\rm model}$ is the flux of our synthetic SEDs, $F^{\log}_{\rm obs}$ is the average flux of the logarithm of the data in a given bin as
\begin{equation}
    F^{\log}_{\rm obs} = {\langle \log F_{\rm obs} \rangle}_{\rm bin},
\end{equation}
and the error considered in each bin reads as
\begin{equation}
    \sigma^{\log}_{\rm obs} = \max(0.1\log F_{\rm obs},\sigma(\log F_{\rm obs})),
\end{equation}
where $\sigma(\log F_{\rm obs})$ represents one standard deviation of the logarithm of the fluxes in a bin.
For each model, we considered all the six lines of sight for which the RT calculation is performed and we take the lowest $\chi^2$.

The results of the $\chi^2$ analysis applied to our 20 models are reported in Table \ref{tab:chisquare_test}.
We find six models with $\chi^2 \leq 1$, which are favoured by this analysis. \textit{Among these six models, five of them have a grain size distribution cut at $\amin=0.1~\mum$.} This is mainly a consequence of the constraints imposed by the optical/UV data, which are best explained by those models that have a grey extinction at these wavelengths. The only marginally favored ($\chi^2 = 1.00$) model without the cut $\amin=0.1~\mum$ has a MW-type intrinsic extinction curve, which produces strong PAHs features, whose presence in high-redshift quasars is still questioned\footnote{The presence of PAHs in AGN hosting galaxies is unlikely both for observational and theoretical reasons. From the observational point of view, Spitzer/IRS spectra of low redshift quasars do not show PAHs features as prominent as in normal star-forming galaxies \citep[e.g.][]{Hernan-Caballero2016MNRAS}. From the theoretical side, these complex, small, fragile molecules may not survive to the intense radiation of AGN, because of Columb explosion \citep{Tazaki:2020Charging} and/or  drift-induced sputtering \citep{Tazaki:2020Drift}.}.

Five of the six models have either a small or intermediate dust-to-metal ratio ($f_{\rm d} = 0.08$ or $f_{\rm d} = 0.15$), suggesting dust masses of $(3.3 - 6.2) \times 10^7~\msun$. In particular, the models with $f_{\rm d} = 0.15$ have the lowest $\chi^2$, indicating that an intermediate dust content provides the best compromise between the attenuation at the shortest wavelengths and re-emission in the FIR. The best model with $f_{\rm d} = 0.3$ has $\chi^2 = 0.89$, so it is only marginally favoured, emphasizing the fact that models with high $f_d$ lead to a stronger attenuation in the optical/UV than suggested by the data, even if a grey extinction curve is assumed.

Overall, we find no strong preference in favor of any specific dust model composition, with three models having an intrinsic SMC-type dust and three models MW-type dust. However, this is mostly due to the fact that the grain size distribution of silicates and carbonaceous grains are quite similar for the SMC and MW dust models if considering only grains with sizes larger than $0.1~\mum$. In particular, we underline that, in the MW models with $a_{\rm min}=0.1~\mum$, PAH molecules are effectively removed. 
All of the six models have the fiducial AGN SED, therefore we conclude that the UV-steep SED is disfavoured by this analysis.

\begin{table}
	\centering
	\begin{tabular*}{0.49\textwidth}{ c c c c c c}
		\hline\noalign{\smallskip}
		AGN SED & $f_{\rm d}$ & Dust model & $\amin$ [$\mum$] & $\chi^2$ \\
		\hline\hline\noalign{\smallskip}
		fiducial & $0.08$ & SMC & -    & 1.10 \\
		fiducial & $0.3$  & SMC & -    & 3.23 \\
		fiducial & $0.08$ & MW  & -    & 1.00 \\
		fiducial & $0.3$  & MW  & -    & 3.99 \\
		UV-steep & $0.08$ & SMC & -    & 1.48 \\
		UV-steep & $0.3$  & SMC & -    & 4.82 \\
		UV-steep & $0.08$ & MW  & -    & 2.34 \\
		UV-steep & $0.3$  & MW  & -    & 7.00 \\
		fiducial & $0.08$ & SMC & 0.01 & 1.06 \\
		fiducial & $0.3$  & SMC & 0.01 & 3.10 \\
		fiducial & $0.08$ & SMC & 0.1  & 0.68 \\
		fiducial & $0.3$  & SMC & 0.1  & 0.97 \\
		fiducial & $0.08$ & MW  & 0.1  & 0.60 \\
		fiducial & $0.3$  & MW  & 0.1  & 1.31 \\
		UV-steep & $0.08$ & SMC & 0.1  & 1.31 \\
		UV-steep & $0.3$  & SMC & 0.1  & 3.17 \\
		UV-steep & $0.08$ & MW  & 0.1  & 1.71 \\
		UV-steep & $0.3$  & MW  & 0.1  & 4.14 \\
		fiducial & $0.15$ & SMC & 0.1  & 0.52 \\
		fiducial & $0.15$ & MW  & 0.1  & 0.57 \\
	\end{tabular*}
  	\caption{The set of 20 models considered in our $\chi^2$ analysis, in order to find the best match with the observed sample. The columns indicate: (first) the AGN intrinsic SED adopted, (second) the dust-to-metal ratio, (third) the dust model as in \citet{Weingartner:2001}, (fourth) the minimum grain size in the modified grain size distribution (if it is not specified, no cut is applied, and the standard grain size distribution is considered), (fifth) the $\chi^2$ for the model.
  	  \label{tab:chisquare_test}
  	  }
\end{table}

\begin{figure*}
	\centering
	\includegraphics[width=0.475\textwidth]{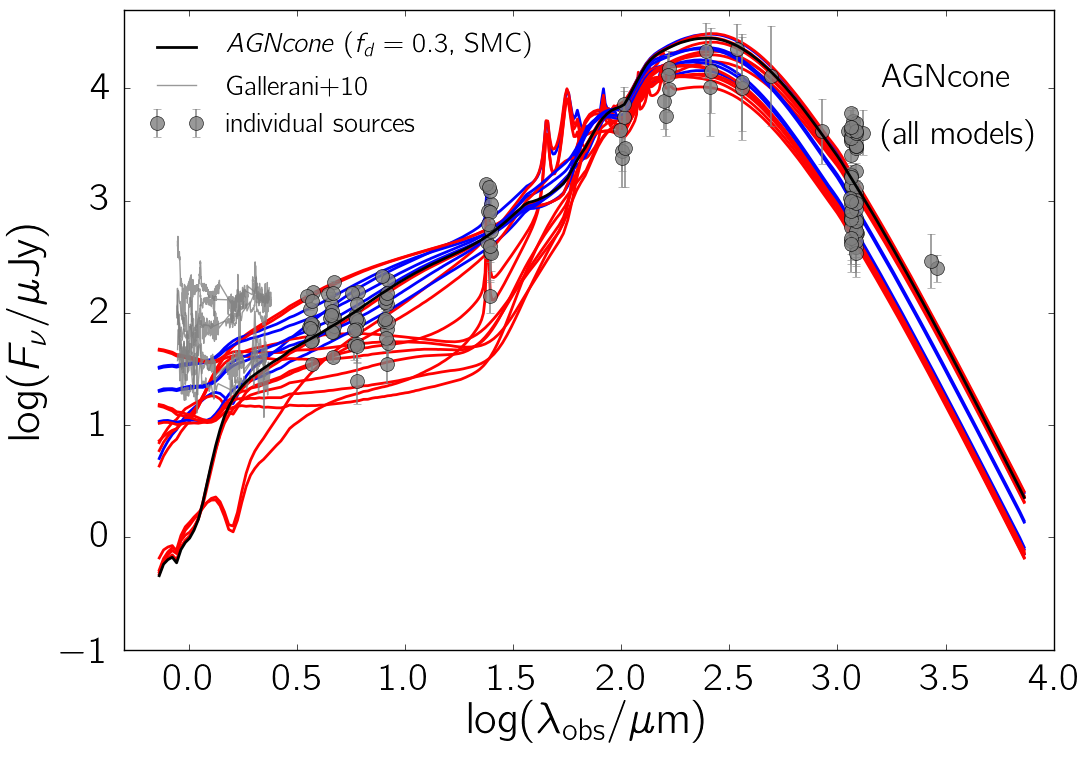}
	\hfill
	\includegraphics[width=0.475\textwidth]{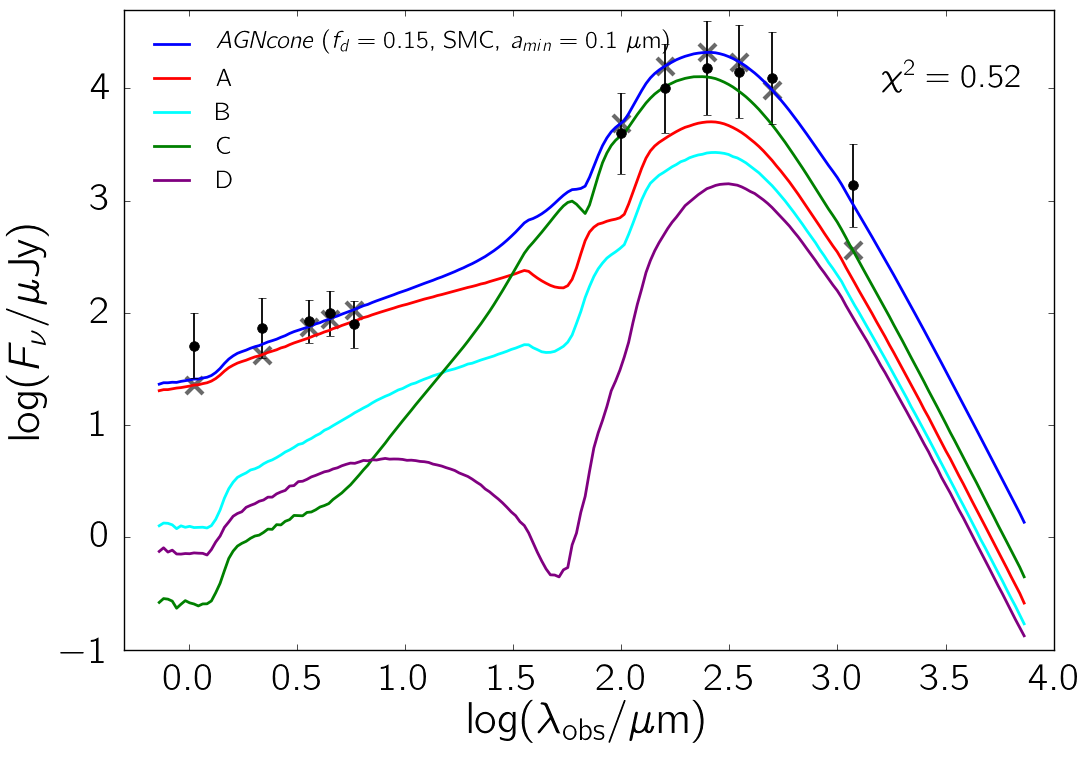}
	\caption{\emph{Left:} Comparison of the \emph{total} SEDs of the 20 models considered in the $\chi^2$-analysis (coloured lines) with UV-to-FIR data. The model $f_{\rm d} = 0.3$, SMC-type dust and the fiducial AGN SED, considered as fiducial in \citetalias{DiMascia:2021}, is plotted with a black solid line. We also color-code each model according to the value of $\chi^2$: models with $\chi^2>1$ are shown in red, models with  $\chi^2 \leq 1$ in blue. The photometric data of the sources considered are shown with grey points, redshifted at the redshift of the simulated snapshot. The optical/UV spectra for the $z>6$ quasars analysed in \citetalias{Gallerani:2010} are shown with grey lines. \emph{Right:} Synthetic SEDs of the \emph{individual} sources for the best fit model, which has $f_{\rm d}=0.15$, fiducial AGN SED, and SMC dust composition modified with $\amin = 0.1~\mum$. The color legend for each source is the same as in Fig. \ref{fig:dust_map}, and the SED of the whole field of view is shown in blue. In particular, the SED of the UV-bright source (A) is plotted in red, while the one of the IR-bright source (C) in green. We mark with grey crosses the fluxes of our synthetic SEDs used for the $\chi^2$ computation. We also show the observed data grouped into eleven bins (see text for more details) with black circles. The $\chi^2$ of the best fit model is reported in the top right corner. 
	    \label{fig:best models}
	}
\end{figure*}

In the right-hand panel of Fig. \ref{fig:best models} we show the best fit model, given by $f_{\rm d}=0.15$, the fiducial AGN SED and the cut in the grain size distribution at $\amin=0.1~\mum$ from an original SMC-type dust composition. The cut in the grain size distribution results in a grey extinction at short wavelengths, and provides the best agreement with the optical/UV data, whereas the intermediate dust content given by $f_{\rm d}=0.15$ (i.e. dust masses of $\approx 6 \times 10^7~\msun$) gives the best compromise between UV attenuation and IR emission. We further comment on the attenuation curves in Section \ref{best_attenuation}. 

\subsection{Attenuation curves in high-redshift quasars} \label{best_attenuation}

The $\chi^2$ analysis in Section \ref{sec:best_model} provides us with important insights concerning the dust abundance, composition and grain size distribution in $z\sim6$ quasars. We can therefore infer an attenuation curve that correctly reproduces the current observations, according to our framework. 

\begin{figure*}
	\centering
	\includegraphics[width=0.475\textwidth]{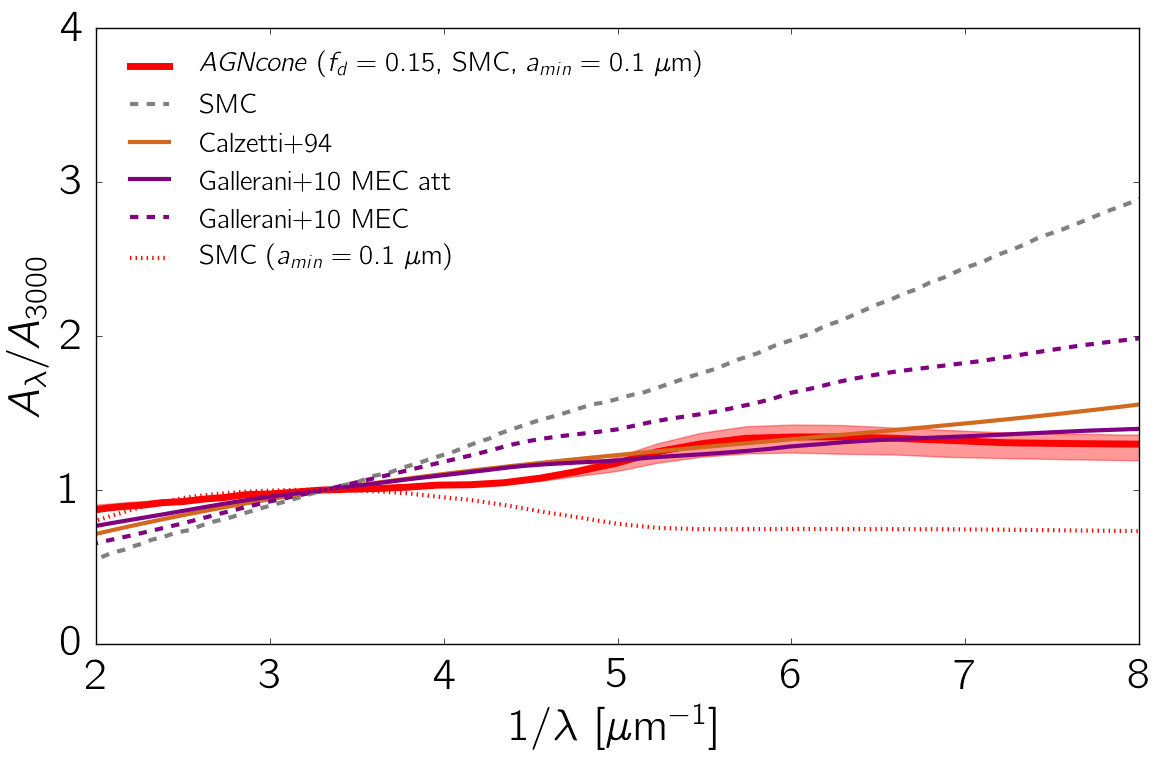}
	\hfill
	\includegraphics[width=0.475\textwidth]{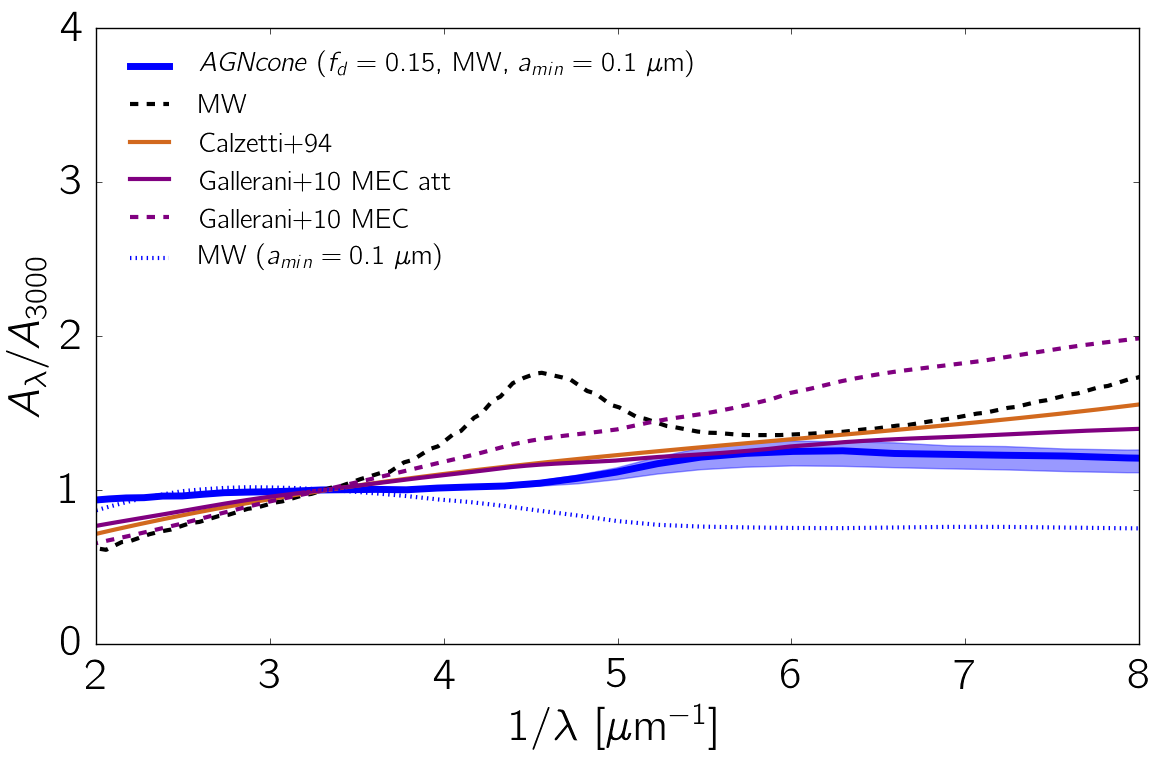}
	\caption{Attenuation curves for the best-fit models, characterised by $f_{\rm d}=0.15$, fiducial AGN SED, $\amin=0.1$ $\mum$ and SMC (left panel) or MW (right panel) original dust composition. As a reference, we also plot the SMC extinction curve (grey dashed line), in the left panel, whereas we show the MW extinction curve (black dashed line) in the right panel. We also plot the Calzetti attenuation curve \citep[light brown solid line,][]{Calzetti:1994}, and the extinction curve (MEC, purple dashed line) and attenuation curve (MEC attenuated, purple solid line) obtained in \citetalias{Gallerani:2010} for high-redshift quasars. Finally, we show the intrinsic extinction curve for the modified grain size distribution used in the SMC (red dotted line) and MW (blue dotted line) case.
	    \label{fig:best models_ext}
	}
\end{figure*}

In the left panel of Fig. \ref{fig:best models_ext} we show the attenuation curve for our best-fit model, which has the fiducial AGN SED, $f_{\rm d}=0.15$ and a SMC-like dust composition, with a grain size distribution cut at $\amin=0.1~\mum$. It displays an almost flat attenuation curve, with small los-to-los variations. Compared to the SMC extinction curve, it shows a large deviation at short wavelengths, where the SMC curve steepens. 
The second best-fit model has the same setup of the best one, except with a MW-like dust composition. However, the results are almost the same, because the modified grain size distribution in these two models result in a very similar dust extinction curve. The corresponding attenuation curve is shown in the right panel of Fig. \ref{fig:best models_ext}. We note that in the MW-derived model the characteristic bump at $2175~\angstrom$ is absent because both the PAH molecules and small graphite grains, which contribute to the feature \citep[e.g.][]{Li2001ApJ554778L}, are removed by the cut at $\amin = 0.1$ $\mum$.
The attenuation curves of these two models predicted by our calculation are slightly flatter than the \citet{Calzetti:1994} one, and they are in excellent agreement with the average attenuation\footnote{We note that \citetalias{Gallerani:2010} refer to their results in terms of {\it extinction} curve. This notation is commonly used in the case of quasars, typically represented as a point source plus a foreground screen geometry. However, in this work we find that the attenuation curve substantially differs from the extinction curve even in the case of AGN (see for example Fig. \ref{fig:attenuation_curves} and Fig. \ref{fig:best models_ext}). This implies that a simplified geometry cannot fairly represent quasars for at least two reasons: 1) emission from stars cannot be neglected in these systems since quasar host galaxies are typically highly star-forming ($SFR\sim 10^2-10^3~\rm M_{\odot}~yr^{-1}$); 2) dust distribution in quasar host galaxies is shaped by quasar feedback and presents a complex, disturbed morphology. In light of these results, the curves found by \citetalias{Gallerani:2010} from their fitting procedure should be interpreted as {\it attenuation} curves, and not as extinction curves.} curve deduced in \citetalias{Gallerani:2010}, with a small deviation only at the shortest wavelengths, where our curves are flatter.

We underline that the attenuation curves of our two best models shown in Fig. \ref{fig:best models_ext} are derived by collecting the radiation of the whole field of view. However, for a fair comparison with the observations of optical/UV-selected quasars, it might be more appropriate to investigate the attenuation curve relative to the brightest UV source in the field of view. We checked that the attenuation curve inferred by considering only the radiation from source A differ by less than $10~\%$ from the attenuation curve derived by considering the whole field of view, shown in in Fig. \ref{fig:best models_ext}.

\section{Discussion} \label{sec:discussion}

\subsection{Implications}

Our calculations support a scenario in which dust in AGN-hosting galaxies is dominated by large grains (i.e. $\amin = 0.1$ $\mum$). This can either be explained by very efficient coagulation of small grains into larger ones in the dense molecular clouds of quasar-hosts \citep[see e.g][]{Nozawa:2015} or by physical mechanisms that preferentially destroy small grains in the extreme environments around AGN. Recently, two processes have been studied in this context. First, dust destruction by charging can wipe out small grains due to the intense radiation field in the AGN proximity \citep{Tazaki:2020Charging}. Second, grains might be destroyed by drift-induced sputtering while they are moved from the equatorial plane toward the polar region by the radiation pressure from the AGN. This mechanism is more effective at destroying small grains than large grains, effectively selecting the dust component that is directly irradiated by the central source and that is responsible of the extinction \citep{Tazaki:2020Drift}.
The two processes might also act at the same time. 

The framework used in this work assumes a linear scaling relation between dust and metals; furthermore, the same grain size distribution is assumed for all the dust in the field of view. Therefore, we cannot discriminate between a scenario in which small grains are absent in most of the ISM of AGN-host galaxies, or if the grain size distribution supported by our analysis reflects only the properties of a dust component located in the proximity of the AGN. 
An accurate description of the dust content and the grain size distribution would require to follow in detail the processes involved in the grains formation and destruction, whose efficiency depends on the local gas physical conditions, such as density, temperature, metallicity \citep[see e.g.][]{Hirashita:2019, Aoyama:2020}. This detailed treatment of the dust physics is beyond the scope of our work.

Moreover, if large grains are dominant because of dust production mechanisms in place at high redshift, this evidence further supports the role of SNe as the main dust factories in the early Universe \citep{Todini:2001}. A SN-origin for dust at high redshift was also suggested by the extinction curve of J1048+4637, measured by \citet{Maiolino:2004}, which flattens at $\lambda > 1700~\angstrom$ and then rises again at $\lambda < 1700~\angstrom$. This evidence suggests that, even though we find attenuation curves in AGN to be flatter than the SMC on average \citepalias[as also found in]{Gallerani:2010}, we cannot exclude the presence of a minor component of small grains, which can be unveiled by studying individually each source. We plan to improve our work in the future in this direction, and to include also models of dust grains from SN origin in our RT calculations. Furthermore, in order to reproduce the extinction curve of J1048+4637 measured by \citet{Maiolino:2004}, \citet{Nozawa:2015} had to assume in their dust models that carbonaceous grains were in the form of amorphous carbon and not graphite, which allowed them to remove the $2175~\angstrom$ bump. We also plan to consider models with amorphous carbon in a future work. 

\subsection{Effective dust-to-metal ratio in presence of efficient dust destruction}

The models favored by our $\chi^2$-analysis in Section \ref{sec:best_model} all lack small grains. In practice, their grain size distribution has the same functional form of the original model used (i.e. SMC or MW) with a cut at $\amin=0.1~\mum$. Instead, the dust mass considered in the simulations for these models, which is inferred from the dust-to-metal ratio (equation \ref{eq:fd}), is the same as for the models with the standard grain size distribution, i.e. it is simply re-distributed among larger grains. 

In this Section, we assume that the dominance of large grains is the result of some destruction process in AGN environments \citep[see e.g.][]{Tazaki:2020Charging, Tazaki:2020Drift}. Under this hypothesis, the effective dust production in the AGN-hosting galaxy is higher than if destruction is not accounted for. Therefore, the \emph{effective} dust-to-metal ratio (i.e. before accounting for the removal of small grains) is actually higher, and this discrepancy depends on the fraction of grains destroyed. 
In the following, we quantify the factor that relates the assumed dust-to-metal ratio with the effective one for the models with $\amin=0.1~\mum$, under the assumption that small grains are destroyed.

\begin{figure}
    \centering
    \vskip\baselineskip
    \includegraphics[width=0.475\textwidth]{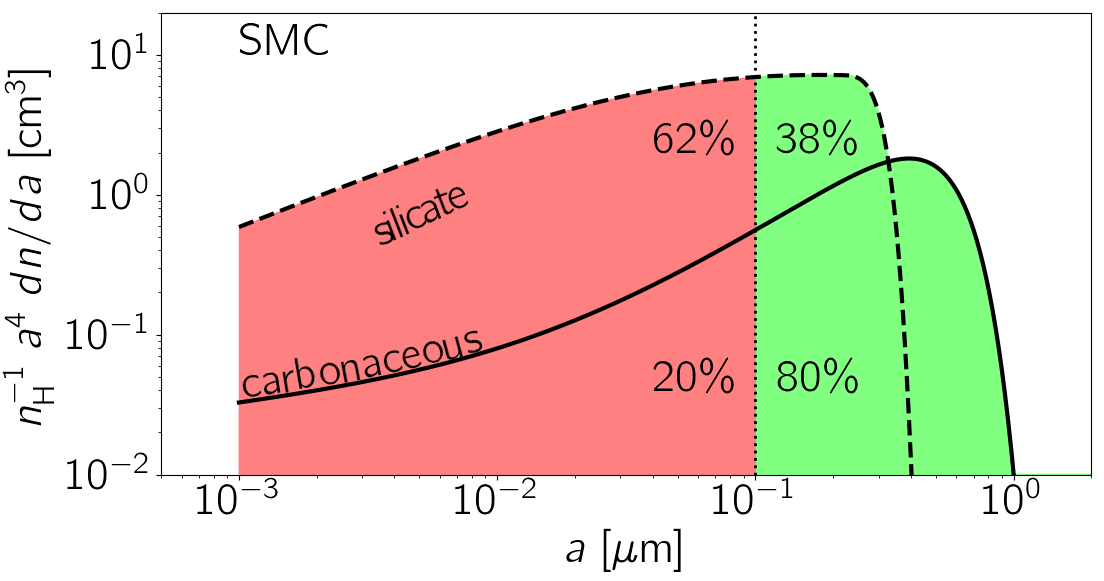}
    \vskip\baselineskip
    \caption{Grain size distributions for the SMC, according to the \citet{Weingartner:2001} model. The solid line refers to carbonaceous grains and the dashed line to silicates. The red regions indicate the grains removed in our models without small grains, i.e. $\amin=0.1~\mum$, while the green regions the grains kept. The percentage quoted refer to the mass fraction of the corresponding grains removed (when the number is in the red region) or kept (green region), for silicate (upper values, close to the dashed line) or carbonaceous (lower values, close to the solid line).
    \label{fig:modified_dnda}
    }
\end{figure}

Given a grain size distribution $dn/da$, which extends from a minimum grain size $\amin$ to a maximum grain size $\amax$, the total dust mass can be expressed as: 
\begin{equation}
    M^{\rm total}_{\rm d} = \frac{M_{\rm gas}}{\overline{\mu}\mp} \int_{\amin}^{\amax} \frac{4}{3}\pi \delta_{\rm g} a^3 \left(\frac{dn}{da}\right) da,
\end{equation}
where $\overline{\mu}$ is the average molecular weight of the gas, $\mp$ is the proton mass, and $\delta_{\rm g}$ represents the bulk density of the grain, which we assume to be $\delta_{\rm carb} = 2.24~\gcc$ for carbonaceous grains and $\delta_{\rm sil} = 3.5~\gcc$ for silicate grains as in \citet{Weingartner:2001}. 
The total mass fraction of grains smaller than $0.1~\mum$ for a fixed species is then given by the ratio:
\begin{align*}
    \frac{M^{\rm small}_{\rm d}}{M^{\rm total}_{\rm d}} = &\frac{\frac{M_{\rm gas}}{\overline{\mu}\mp}\int_{\amin}^{0.1\mum} \frac{4}{3}\pi \delta_{\rm g} a^3 \left(\frac{dn}{da}\right) da}{\frac{M_{\rm gas}}{\overline{\mu}\mp}\int_{\amin}^{\amax} \frac{4}{3}\pi \delta_{\rm g} a^3 \left(\frac{dn}{da}\right) da} = \frac{\int_{\amin}^{0.1\mum} a^3 \left(\frac{dn}{da}\right) da}{\int_{\amin}^{\amax} a^3 \left(\frac{dn}{da}\right) da} = \\ 
    &= \frac{\int_{\amin}^{0.1\mum} a^3 \left(\frac{dn}{da}\right) da}{\int_{\amin}^{0.1\mum} a^3 \left(\frac{dn}{da}\right) da + \int_{0.1\mum}^{\amax} a^3 \left(\frac{dn}{da}\right) da} = \\ &= \frac{\Rsmall}{\Rsmall + \Rlarge} = \frac{\Rsmall}{\Rtotal},
\end{align*}
where we introduced the \emph{reduced mass} of small and large grains $\Rsmall$ and $\Rlarge$ respectively, which do not depend on the bulk density. We also indicate $\Rtotal = \Rsmall + \Rlarge$ for a single species.
In \code{SKIRT} the SMC and MW dust models are implemented according to the model by \citet{Weingartner:2001}\footnote{Grain size distribution as in equations 4, 5 and 6, coefficients from their Table 1 and Table 3.}. By using the appropriate values for the SMC model, we get:
\begin{align}
    \frac{\mathcal{R_{\rm small}^{\rm SMC, sil}}}{\mathcal{R_{\rm total}^{\rm SMC, sil}}}  &= 0.62, \\
    \frac{\mathcal{R_{\rm small}^{\rm SMC, carb}}}{\mathcal{R_{\rm total}^{\rm SMC, carb}}} &= 0.20.
\end{align}
Thus, in the SMC models without grains $\lesssim 0.1~\mum$, $62\%$ of the dust mass in silicates and $20\%$ of the dust mass in carbonaceous grains of the original grain size distribution is removed. These results are summarised in Fig. \ref{fig:modified_dnda}. Overall, the amount of dust mass in small grains (i.e. removed) is:
\begin{equation}
    \frac{M^{\rm SMC, small}_{\rm d}}{M^{\rm SMC, total}_{\rm d}} = \frac{\delta_{\rm sil} \ \mathcal{R_{\rm small}^{\rm SMC, sil}} + \delta_{\rm carb} \ \mathcal{R_{\rm small}^{\rm SMC, carb}} }{\delta_{\rm sil} \ \mathcal{R_{\rm total}^{\rm SMC, sil}} + \delta_{\rm carb} \ \mathcal{R_{\rm total}^{\rm SMC, carb}} } = 0.59. 
\end{equation}
Therefore, the overall dust mass formed, i.e. without considering destruction of small grains, is a factor $1/(1-0.59) \approx 2.5$ higher. This implies that, for a given value of the dust-to-metal ratio $f_{\rm d}$, an SMC model without small grains has an effective dust-to-metal ratio of $f_{\rm d}^{\rm eff} \approx 2.5 f_{\rm d}$. For the best-fitting model found in Section \ref{sec:best_model}, which has $f_{\rm d} = 0.15$, we derive $f_{\rm d}^{\rm eff} = 0.38$, which is very similar to the Milky-Way value. This might indicate that dust production at high-redshift was already quite efficient in AGN-hosting galaxies. 
The calculation illustrated above also implies that there is a maximum dust-to-metal ratio for which we can assume the existence of a physical process responsible for the selective removal of grains smaller than $0.1~\mum$, which is $f^{\rm max}_{\rm d} = 0.4$.

\subsection{AGN feedback and dust spatial distribution}\label{sec:dust_spatial_disc}

In Section \ref{sec:ext_curves}, we studied how the shape of the attenuation curves of a simulated galaxy depends, among many other factors, on the dust-sources geometry. In particular, we find that the attenuation curves in AGN-host galaxies tend to be steeper than in normal star-forming galaxies because of lower dust densities. We argue that this behaviour is a consequence of the AGN feedback, which drives powerful outflows, pushing away the gas from the BH surrounding and outside the galaxies, thus reducing the overall gas density. 
This finding motivates the need for assuming a peculiar grain size distribution in order to reproduce the grey attenuation curves suggested by optical/UV data. We are able to match the observations if grains smaller than $\amin=0.1~\mum$ are removed (see Section \ref{sec:best_model}). 

However, another possibility to ease the tension with the data is that the dust distribution in observed quasars is actually much more compact than what predicted by the hydro-simulations considered in this work. For example, it is possible that the kinetic feedback implemented is too effective in powering winds, and that other prescriptions, such as thermal feedback, might instead be better suited to reproduce the observations. If that is the case, then the higher density would also result in higher dust attenuation. Therefore, in order to investigate this point, we consider the absolute dust attenuation at $3000~\angstrom$, $A_{3000}$, of our best-model and we compare its value with the observations. 
The six quasars with optical/UV measurements used in Section \ref{sec:best_model} have $A_{3000}=0.82-2.00$ \citepalias{Gallerani:2010}. Our best-fit model has $A_{3000}=1.32$, which is nicely consistent with these results. Dust surface densities an order of magnitude higher (required to mimic a grey extinction curve in \AGNcone{}, see Section \ref{sec:ext_curves}) would result in an attenuation of a factor of $\approx 10$ higher than what predicted, which is too high with respect to the observations. Furthermore, with a higher dust attenuation, a higher intrinsic magnitude would also be required in order to match the far-UV flux of observations. Given that in our framework the bolometric luminosity of an AGN is directly linked with its accretion rate (see Section \ref{Numerical_methods}), this would likely imply super-Eddington accretion. Therefore, we argue that the gas distribution predicted by the hydrodynamic simulations adopted is consistent with observations. Nevertheless, we plan to investigate attenuation curves in simulated galaxies with different feedback prescriptions in a future work.

\subsection{Caveats}

In this work, we consider the snapshot at $z=6.3$ of the run \AGNcone{} performed in \citetalias{paramita:2018}, and we post-process it with RT calculations in order to compare it with the observations of $z\sim 6$ quasars. It is natural to question how representative the chosen DM halo and the specific snapshot we selected are of the quasars sample we compare with. In the low resolution, DM-only simulation performed in \citetalias{paramita:2018} the selected DM halo has a mass of $M_{\rm halo} = 4.4 \times 10^{12}~\msun$ at $z=6$, which is compatible with results from clustering studies \citep[e.g.][]{allevato2016}. However, this DM halo experiences a specific merging and accretion history and cannot be representative of the whole population of high-redshift quasars. Moreover, the evolution of the halo and its host galaxies is also affected by the numerical model implemented, in particular on the AGN feedback, the subgrid physics adopted and the resolution limits. As such, our results might be sensitive to the structure of the simulated halo, and to the properties down to ISM scales, which are ultimately affected by the details of the numerical models implemented. It would be valuable to repeat this analysis by exploring several realizations of the Universe via cosmological hydrodynamical simulations, by considering a wide range of DM halo masses, histories, and the parameters describing the AGN feedback and the subgrid physics. This is beyond the scope of this work.

Moreover, in the analysis performed in Section \ref{sec:best_model}, we compare our simulated AGN with the average properties of the $z \sim 6$ quasars. It would be valuable to also perform this analysis by considering the observed quasars individually, and finding for each of them a simulated object that best matches their SEDs. However, there is only handful of objects with a SED sampled well enough for such calculation. Furthermore, a very high number of simulations would be required for this kind of study.

The snapshot at $z=6.3$ we focus on was already analysed in \citetalias{DiMascia:2021}, where we show it matches reasonably well the average properties of high-redshift quasars, underestimating only the UV emission. In particular, this system is consistent with recent observations of quasars-SMGs companions \citep[e.g.][]{Decarli:2017}. However, three AGN are active (see Section \ref{radiation_sources}) at this snapshot, thus it might represent a peculiar moment in the DM halo evolution. As a check, we post-process additional snapshots (see Appendix \ref{appendix:add_runs}), when a different number of AGN is active. We find that in all cases the dust-attenuation in the UV make our simulated objects under-luminous with respect to the observations, confirming that the need of less extinction at short wavelength is not special of the snapshot selected for the detailed analysis in this work.

As discussed in Section \ref{hydro_sim}, the simulations considered in this work achieve a maximum spatial resolution of $\approx 200$~pc, and a maximum density of $\approx 200~\cc$. Therefore, we are not able to resolve dense, clumpy molecular structures on $\sim$~pc scales, which might dominate the extinction with respect to the diffuse gas in the ISM. Even the highest resolution cosmological simulations do not reach sub-pc scale-resolution, i.e. cannot follow in detail all the physical processes inside such clumpy molecular clouds. Resolving such structures might decrease the effective opacity of the medium, because of their low volume filling factor, as shown in radiative transfer simulations in two-phases clumpy media \citep[e.g.][]{Witt96, Bianchi00}. This might in turn imply a higher dust mass than suggested by our simulations in order to match the observational data. However, this effect is highly dependent on the ISM properties, for example \citet{Decataldo2020} find that only $10\%$ of the UV flux can escape such clumpy structures. While dedicated works will be needed to fully explore the subject, the predicted attenuation curve steepening should only be marginally affected by such differential variation. Thus we expect our main conclusion that small grains are removed in AGN environment to still hold. 

\section{Summary and conclusions} \label{sec:conclusions}

We have studied dust attenuation in $z\sim6$ quasar-host galaxies by post-processing with radiative transfer (RT) calculations a suite of cosmological zoom-in hydrodynamical simulations. Our main goal has been to clarify whether extinction curves such as those found in the the Small Magellanic Cloud (SMC) or in the Milky Way (MW) can explain multi-wavelength observations of high-redshift quasars spectra. 
Using the cosmological simulations presented in \citetalias{paramita:2018}, in which the evolution of a $\sim 10^{12}~\msun$ dark matter halo at $z\simeq 6$ is followed in a zoom-in fashion, we focus on the snapshot at $z=6.3$ of the run \AGNcone{}, previously analyzed in detail in \citetalias{DiMascia:2021}. At this evolutionary stage, the simulated halo hosts three galaxies with active black holes at their centres, corresponding to an (unattenuated) UV magnitude of $M_{\rm UV} = -28$. 
We have post-processed the snapshot with the code \code{SKIRT} \citep{Baes:2003, Baes:2015, Camps:2015, Camps2016}, which solves the radiative transfer problem in dusty systems. We considered different models for the overall dust mass content, dust composition, grain size distribution, and intrinsic AGN spectrum. 
We have first examined the effects of relative dust-sources geometry on the synthetic attenuation curves by comparing the results of \AGNcone{} with a control run without BHs, namely \noAGN{}. We find that, at fixed dust composition, \noAGN{} shows flatter attenuation curves and smaller los-to-los variations than \AGNcone{} (Fig. \ref{fig:attenuation_curves}).

We quantify the steepness of the attenuation curves by the ratio $A_{0.1}/A_1$, which compares the total attenuation at $0.1~\mum$ and $1~\mum$ respectively. We provide an analytical model that relates this ratio with the V-band optical depth, for two simplified geometries: i) a point source surrounded by a spherical distribution of dust (\emph{point source model}) and ii) a sphere with dust and emitters uniformly distributed (\emph{intermixed model}). In both cases, the ratio $A_{0.1}/A_1$ decreases for $\tauV \gg 1$, and in the intermixed model it goes to $1$, thereby resulting in a flat attenuation curve. 
The sphere model well describes the compact dust/stars distribution in \noAGN{} run, and the low $A_{0.1}/A_1$ ratio predicted at high optical depths is consistent with the most (intrinsic) UV-emitting regions having $\tauV > 10$ in this run (Fig. \ref{fig:attenuation_vs_optical_depth}, right panel). In the \AGNcone{} run, they are characterised by $1 < \tauV < 10$, and the value of the $A_{0.1}/A_1$ ratio lies in the middle of the results of the point source and intermixed models. Overall, the fact that the attenuation curve in $\noAGN{}$ is flatter than in \AGNcone{} is well explained by the higher dust optical depths.
The different optical depths, i.e. dust distribution, among the two runs is a consequence of the AGN feedback, which drives powerful outflows that remove gas from the central regions, and distribute it over several kpc. We conclude that the steep synthetic attenuation curves predicted for the simulated AGN-host galaxies in \AGNcone{} are a direct consequence of a gas distribution affected by AGN activity.

We then compare the synthetic Spectral Energy Distributions (SEDs) obtained by RT post-processing \AGNcone{} with multi-wavelength observations of a sample of bright $z\sim6$ quasars with comparable UV magnitudes (${-29\lesssim M_{\rm UV}\lesssim -26}$). Among the twenty models considered, only six are favoured by a $\chi^2$-analysis (Table \ref{tab:chisquare_test} and Figure \ref{fig:best models}). Five of these six models have a modified grain size distribution, obtained from the one underlying the original dust model (SMC-like or MW-like) by removing grains with sizes $a < \amin=0.1~\mum$. The attenuation curves inferred for these models are close to flat (Fig. \ref{fig:best models_ext}). 
These findings are consistent with the results by \citetalias{Gallerani:2010}, who fit the spectra of a sample of $3.9 \leq z \leq 6.4$ reddened quasars, suggesting extinction curves flatter than the SMC. 

The standard dust models tend to produce attenuation curves that are too steep to match the data, because of the low optical depths in AGN-host galaxies caused by AGN activity. A modified grain size distribution is needed in order to reconcile the synthetic SEDs with the optical/UV data. Therefore we caution toward the applicability of the SMC extinction law in high-$z$ quasar-hosts when interpreting observations.

Our calculations finally suggest a dust-to-metal ratio $f_{\rm d} \lesssim 0.15$, which implies dust masses $M_{\rm d} \lesssim 6 \times 10^7~\msun$ for the AGN-hosts in the sample. However, if we attribute the modified grain size distribution to efficient destruction processes in place in AGN-environments \citep[e.g.][]{Tazaki:2020Drift,Tazaki:2020Charging}, the original dust mass produced might instead be a factor $\approx 2.5$ higher, thus implying an effective dust-to-metal ratio of $\approx 0.4$ before accounting for dust removal (Fig. \ref{fig:modified_dnda}). This would suggest very efficient dust production in high-redshift AGN-hosting galaxies. 
Alternatively, the dominance of large grains in AGN-hosts might suggest a supernova origin for dust at high redshift \citep[e.g.][]{Todini:2001} or efficient coagulation of small grains into larger ones \citep[see e.g][]{Nozawa:2015}.
Which of the two scenarios applies, i.e. efficient  destruction of small grains vs. preferential production of large grains, cannot be assessed from our calculations, because of limited spatial resolution and the lack of detailed physics in the hydrodynamical simulations. A framework describing the dust content and grain size evolution according to the local gas physical conditions (e.g. density, temperature, metallicity) in the galaxy, paired with a higher spatial resolution, is required to disentangle the two scenarios. These improvements are left for a future work. 

The upcoming James Webb Space Telescope (JWST) will significantly improve current rest-frame optical/NIR data with deeper observations and it will provide us with high-resolution spectra, effectively probing the spectral region more sensitive to dust attenuation with an unprecedented sensitivity. This will drive a significant contribution in our understanding of the dust origin in the early Universe.

\section*{acknowledgements}
The authors thank the anonymous referee for carefully reading the manuscript and for the useful comments, which improved the quality of the work. FD thanks Laura Sommovigo for helpful discussions and Christoph Behrens for code support. SG acknowledges support from the ASI-INAF n. 2018-31-HH.0 grant and PRIN-MIUR 2017 (PI Fabrizio Fiore). We acknowledge usage of the Python programming language \citep{python2,python3}, Astropy \citep{astropy}, Cython \citep{cython}, Matplotlib \citep{matplotlib}, NumPy \citep{numpy}, \code{pynbody} \citep{pynbody}, and SciPy \citep{scipy}.

\section*{Data availability}
Part of the data underlying this article were accessed from the computational resources available to the Cosmology Group at Scuola Normale Superiore, Pisa (IT). The derived data generated in this research will be shared on reasonable request to the corresponding author.

\bibliographystyle{mnras}
\bibliography{file_bibliography/ref}

\appendix

\section{Additional runs} \label{appendix:add_runs}

The analysis performed in this work is focused on the snapshot at $z=6.3$. In order to check that the conclusion we derive are not sensitive to this particular choice, we also post-process the snapshots at $z=6.2, 6.7, 7$, exploring moments in the halo history, characterised by a different AGN activity. In particular: at $z=7$ a single AGN with $\dot{M}_{\rm BH} = 7~\msunyr$ is present; at $z=6.7$, an AGN with $\dot{M}_{\rm BH} = 16~\msunyr$ is in place, with other two much less active ones with $\dot{M}_{\rm BH} \lesssim 2~\msunyr$; at $z=6.2$, a very powerful AGN with $\dot{M}_{\rm BH} = 68~\msunyr$ dominate the emission, with other two AGN accreting at $\dot{M}_{\rm BH} \lesssim 2~\msunyr$. In each of these snapshots the intrinsic UV emission is provided mainly by a single source.

\begin{figure*}
    \centering
    \hfill
    \includegraphics[width=0.475\textwidth]{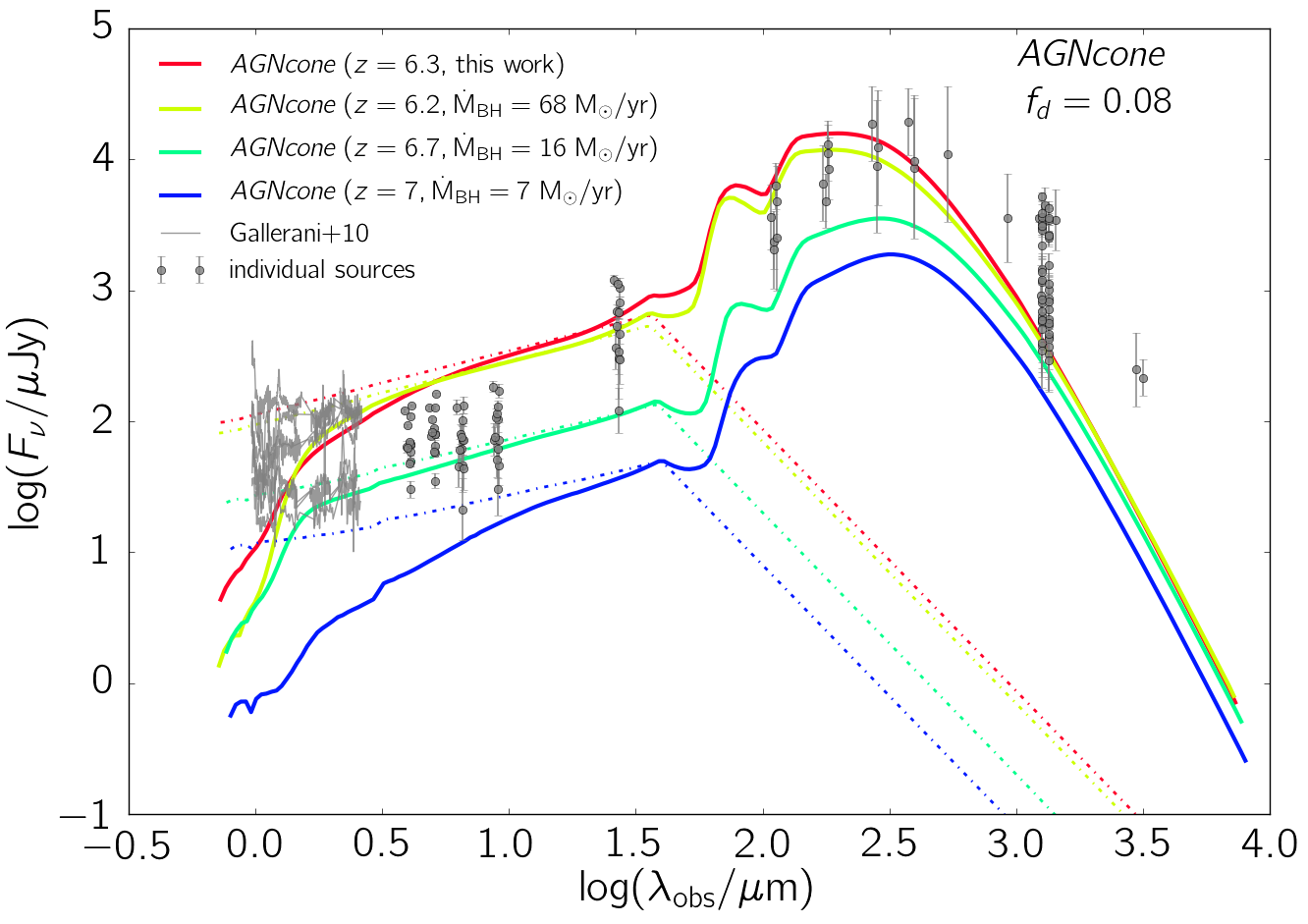}
    \hfill
    \includegraphics[width=0.475\textwidth]{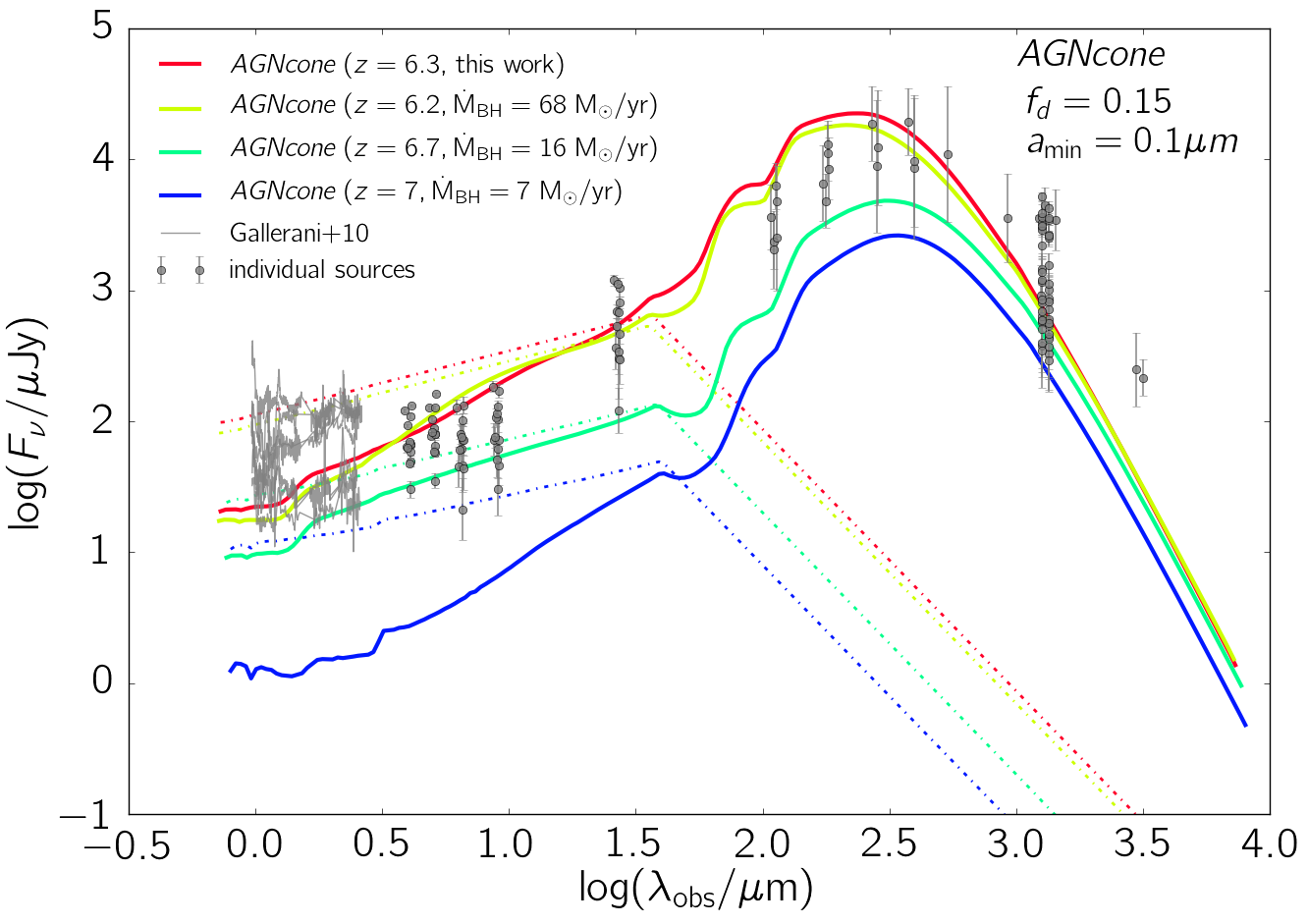}
    \caption{Comparison of the predicted SEDs with quasar observations for the snapshots at $z=6.2, 6.3, 6.7, 7$ (with $z=6.3$ being the snapshot analyzed in this work). The left panel shows the result obtained with a dust model with $f_{\rm d}=0.08$ and SMC extinction curve. The right panel refers to the best-fit model, characterised by $f_{\rm d}=0.15$, SMC extinction curve and minimum grain size $\amin = 0.1~\mum$. The quasar data shown refer to the sources in Table \ref{tab:sources_comparison}.
    \label{fig:RT_add_snapshots}
    }
\end{figure*}

In Fig. \ref{fig:RT_add_snapshots} we show the SEDs at these snapshots for a dust model with $f_{\rm d}=0.08$ and an SMC extinction curve (left panel) and for the best-fit model found in Section \ref{sec:best_model} (right panel). We find that the SEDs produced with the former dust model fail to reproduce the UV data, as for the snapshot at $z=6.3$. Instead, the discrepancy with the observations is significantly reduced for the best-fit model.

The models that do not agree well with the data have a lower intrinsic bolometric luminosity with respect to the observations, suggesting that the mismatch is due to the fact that the observed quasars are probably in a more active phase with respect to the AGN in the simulation at these snapshots.

This analysis confirms that the need of less extinction at short wavelength is not special of the snapshot at $z=6.3$.

\bsp
\label{lastpage}

\end{document}